\documentclass[aps,preprint,tightenlines,showkeys,nofootinbib]{revtex4}
\usepackage{epsfig,amsmath,amssymb,graphicx,ulem}

\newcommand{\pr}{\mbox{pr}}
\newcommand{\as}{\mathbf{a}}
\newcommand{\ares}{\mathbf{a}_{res}}
\newcommand{\amarg}{\mathbf{a}_{marg}}
\newcommand{\bs}{\mathbf{b}}
\newcommand{\bres}{\mathbf{b}_{res}}
\newcommand{\bmarg}{\mathbf{b}_{marg}}

\begin{document}
\title{Bayesian Methods for Parameter Estimation in Effective Field Theories}
\author{M.~R.~Schindler}\email{schindle@ohio.edu}
\affiliation{Department of Physics and Astronomy, 
Ohio University, Athens, OH 45701, USA\\}
\author{D.~R.~Phillips}\email{phillips@phy.ohiou.edu}
\affiliation{Department of Physics and Astronomy,
Ohio University, Athens, OH 45701, USA\footnote{Permanent address}\\
and\\
School of Physics and Astronomy, University of Manchester, Manchester, M13 9PL, UK\\}

\date{August 12, 2009}
\begin{abstract}
We demonstrate and explicate Bayesian methods for fitting the parameters that encode the impact of short-distance physics on observables in effective field theories (EFTs). We use Bayes' theorem together with the principle of maximum entropy to account for the prior information that these parameters should be natural, i.e. ${\cal O}(1)$ in appropriate units. Marginalization can then be employed to integrate the resulting probability density function (pdf) over the EFT parameters that are not of specific interest in the fit.
We also explore marginalization over the order of the EFT calculation, $M$, and over the variable, $R$, that encodes the inherent ambiguity in the notion that these parameters are ${\cal O}(1)$. This results in a very general formula for the pdf of the EFT parameters of interest given a data set, $D$. We use this formula and the simpler ``augmented $\chi^2$" in a toy problem for which we generate pseudo-data. These Bayesian methods, when used in combination with the ``naturalness prior", facilitate reliable extractions of EFT parameters in cases where $\chi^2$ methods are ambiguous at best. We also examine the problem of extracting the nucleon mass in the chiral limit, $M_0$, and the nucleon sigma term, from pseudo-data on the nucleon mass as a function of the pion mass. We find that Bayesian techniques can provide reliable information on $M_0$, even if some of the data points used for the extraction lie outside the region of applicability of the EFT.
\end{abstract}


\maketitle

\section{Introduction}\label{Sec:Intro}

Effective field theory (EFT) methods allow the treatment of problems
in which there is a separation of scales. In these theories dynamics
at the low-energy scale, $m$, say, is incorporated explicitly in the
theory, while the degrees of freedom that enter the problem at the
high-energy scale, $\Lambda$, are integrated out. (See, Refs.~\cite{Georgi:1994qn,Manohar95,Phillips02,Kaplan05} for 
pedagogical introductions to EFT.)
The impact of modes
with $p \sim \Lambda$ on dynamics for $p \sim m$ is then accounted for
via a sequence of contact operators of increasing dimension. If there is no
pre-determination
as to which operators appear in this
sequence 
then the theory is free of model assumptions about the high-energy dynamics. Therefore, in general,
all contact operators consistent with the symmetries that are
applicable at the scale $p \sim m$ should be included in the EFT expansion.
The coefficients of these admissible contact operators encode the impact of
high-energy physics on low-energy observables in a systematic and
model-independent way. Observables corresponding to momenta $p \sim
m$ can be computed as an expansion in powers of $m/\Lambda$, and the
resultant formulae are model-independent predictions, depending only
on the existence of the scale separation and the symmetries of the
low-energy theory.

One popular application of EFT is to low-energy QCD. In this case the
scale separation is between the mass of the pion, the pseudo-Goldstone
boson of QCD's spontaneously-broken (approximate) chiral symmetry, and
the masses of other hadronic degrees of freedom. The EFT which
incorporates chiral symmetry and encodes this scale separation is
known as chiral perturbation theory
($\chi$PT)~\cite{Weinberg79,GL82,Gasser:1983yg,Gasser:1984gg,Scherer:2002tk,BM07,Bernard08}. 
The $\chi$PT expansion for a hadronic observable is then an expansion in powers of $m/\Lambda$, with 
loop diagrams introducing non-analytic dependence on this expansion parameter.\footnote{Here we are concentrating on an observable at a particular kinematic point. $\chi$PT can, of course, also be used to compute the low-energy dependence of observables on $\xi/\Lambda$, where $\xi$ is any kinematic parameter with dimensions of mass.} The dynamics at scale $\Lambda$ impacts this expansion through certain coefficients which are not determined {\it a priori}. The low-energy symmetries of QCD mandate that once determined in one process these parameters---the ``low-energy constants" (LECs) of $\chi$PT---will appear in other processes too, thereby giving $\chi$PT predictive power once the LECs at a given order are known.\footnote{Note that here and throughout we are using the term ``low-energy constants" to refer to the coefficients in the EFT expansion of a physical observable. This is different to the oft-employed meaning of ``LEC" as the coefficient of an operator in the EFT Lagrangian. We have chosen not to adopt that meaning here since the values of those coefficients depend on the conventions used in the Lagrangian: the interpolating fields, set of independent operators written down at a given order, etc. In contrast the coefficients in the EFT expansion of S-matrix elements are independent of all such choices.} There are some instances in which an LEC can be rigorously computed from the underlying theory, but lattice calculations which do this for low-energy QCD exist in only a very few cases. In this situation the only model-independent way to find the LECs is to fit them to experimental data. Such parameter estimation is thus a crucial component of $\chi$PT, and indeed of all EFT programs.

The standard method of determining LECs from data is to perform a fit
using the EFT expansion of that physical quantity at a fixed
order, employing techniques such as least squares or maximum
likelihood. But here we face several dilemmas as regards the``best" way to obtain
the LECs, including:
\begin{enumerate} 
\item Which data should be used to determine the
LEC?  More data is available as the maximum energy of the data set is
increased, but the reliability of a fixed order EFT calculation
decreases as the energy is increased.  
\item What order of EFT
calculation should be used to extract the LEC?  The first one at which
that LEC appears, or the highest one to which the expansion has been
computed?  
\item How should prior constraints on LECs (e.g. from the
requirement of ``naturalness" with respect to the scale $\Lambda$, or
from other processes) be incorporated into the fit?  
\end{enumerate}
In an ideal situation none of these dilemmas matter, and all fitting
paths lead to the same LEC (within errors). But if
only somewhat imprecise experimental data is available in the region
of validity of the EFT then the extracted LEC can be significantly
sensitive to the manner in which the fit is done.

In this paper we argue that Bayesian methods (see e.g.~Refs.~\cite{Sivia,Jaynes:2003}) are ideal for parameter
estimation of LECs in EFTs, and that they resolve all the above
dilemmas. In the Bayesian approach the central object is the posterior
probability distribution function (pdf) for the LECs of interest, say $a_0$ and $a_1$, and we want their
joint, conditional distribution given a data set $D$: ${\rm pr}(a_0,a_1|D)$.
Bayes' theorem gives us the following relation between this and the
more-usually computed ${\rm pr}(D|a_0,a_1)$:
\begin{equation}
{\rm pr}(a_0,a_1|D)=\frac{{\rm pr}(D|a_0,a_1) {\rm pr}(a_0,a_1)}{{\rm pr}(D)}.  
\end{equation}
Here the first factor on the right-hand side is the ``likelihood" that is minimized in a
$\chi^2$ or least-squares approach. It is through the second factor that prior information can be incorporated in the fit. (The
factor in the denominator may be determined by the requirement of a
normalized pdf for ${\rm pr}(a_0,a_1|D)$.)

The fact that EFTs intrinsically depend on scale separation means that
in an EFT fit there is information available on the size of LECs prior
to the analysis of the data. In a standard, perturbative EFT with one
high-energy scale the LECs should be ``natural'' with respect to the
scale $\Lambda$, i.e. ${\cal O}(1)$ when measured in units of
$\Lambda$.\footnote{The situation can be complicated by the presence
of additional scales~\cite{Cohen96,PP03}, and/or infra-red fixed points~\cite{vanKolck99,KSW98A,BMcGR99} but
the underlying fact of prior information on the LECs remains
true.} Consequently we begin by encoding the fact that the LECs
$a_0, \ldots, a_{M}$ should be natural through the prior ${\rm pr}({\bf a})$. We choose a $M+1$-dimensional Gaussian prior, of width $R$, since that is the
least informed prior if we know the expectation value of $\sum_i
a_i^2$~\cite{Gull:1988}. This yields a version of the ``constrained
curve fitting" method recently advocated for lattice QCD data by
Lepage~\cite{Lepage:2001ym}, Morningstar~\cite{Morningstar:2001je} and others (see e.g.~Refs.~\cite{Draper:2003cg,Chen:2004gp}).

Constrained curve fitting therefore amounts to the computation of a posterior pdf with a Gaussian prior at fixed order $M$. We can eliminate sensitivity to the precise value of $R$ by averaging the resulting pdf over a range of $R$
values, thereby incorporating in our result the inherent ambiguity in 
the notion of ``${\cal O}(1)$" LECs. ``Marginalizing" in this way over unwanted
parameters is a technique used to obtain posterior pdfs that incorporate
the uncertainty that results from systematic differences in the parameters which are
obtained when fits are done in different ways. Therefore we also marginalize
over the order $M$ of the EFT calculation. This necessitates additional
marginalization over the LECs that appear at orders $M > 1$ and so amount to ``nuisance parameters" in our effort to extract $a_0$ and $a_1$. Our final formula for the posterior pdf ${\rm pr}(a_0,a_1|D)$, Eq.~(\ref{Ex:Final:pdf}), therefore involves a sum over $M$, as well as integrals over $R$ and $a_2, \ldots$. The resulting central values and uncertainties for $a_0$ and $a_1$ incorporate a rigorous accounting of the theoretical uncertainty present in a low-order EFT fit.

In Section~\ref{Sec:Bayes} we review Bayesian methods as well as the standard maximum-likelihood technique, derive the formula for the ``augmented $\chi^2$"---a $\chi^2$ which penalizes unnatural values of the fit parameters---and derive our final formula---see Eq.~(\ref{Ex:Final:pdf})---for ${\rm pr}(a_0,a_1|D)$. In Sec.~\ref{Sec:Toy} we apply both the augmented $\chi^2$ and the formula (\ref{Ex:Final:pdf}) to the ``toy" problem of extracting the
coefficients of a power-series approximant to the function
$g(x)=(\frac{1}{2} + \tan(\pi/2 x))^2$  from pseudo-data that is statistically
distributed around the curve $g$. We show that these two Bayesian methods provide an extraction which does not depend on the interval over which
the fit is performed. We also show how they avoid certain
ambiguities that are present in $\chi^2$-minimization. And we find
that reliable information on power-series coefficients can be
gleaned from the data even in cases where standard techniques are powerless.

In Sec.~\ref{Sec:Nucleonmass} we generate pseudo-data with small errors from
the $\chi$PT function for the nucleon mass as a function of the pion
mass. We show that methods based on Eq.~(\ref{Ex:Final:pdf}) are capable of determining the
nucleon mass in the chiral limit, $M_0$, from pseudo-data in the pion-mass
range $m=200$--$500$ MeV. We follow this in Sec.~\ref{Sec:Model} with a similar analysis of pseudo-data
generated from
an underlying function $M_N(m)$ that deviates from the $\chi$PT form above 500 MeV.
Bayesian extractions of $M_0$ from such pseudo-data generated in different ranges of $m$ yield consistent results. This conclusion holds even for fit windows that extend significantly above the pion mass where the $\chi$PT from for $M_N(m)$ ceases to be valid.

These three problems are variants of polynomial regression using
Bayesian methods, a problem that has received extensive treatment in
the literature. A Gaussian prior for the polynomial coefficients is a
common choice~\cite{Young77,BlightOtt75}. However, many
authors~\cite{Young77,Deaton80} also incorporate the `commonly held belief
that the [coefficients] will tend to decrease in absolute value as the
order increases'~\cite{Young77}. The idea of marginalizing over the order of the polynomial fit is also
not new. It is discussed extensively in studies of ``Bayesian Model
Averaging", e.g. Refs.~\cite{Guttman05,Rafferty97}. In EFT applications the basis for regression includes non-polynomial functions and, as will be discussed
extensively in Sec.~\ref{Sec:Toy}, the fit form only has a limited region of applicability. These peculiarities make the problem of parameter estimation in EFTs one that has---as far as we can tell---evaded treatment in the Bayesian literature until now. 

On the EFT side Bayesian methods have been applied to chiral extrapolations of quenched lattice data on the nucleon mass as a function of the pion mass in Ref.~\cite{Chen:2003im}. In this work a technique based on using priors obtained from a subset of the lattice data in the analysis of the full data set was employed. Such iterative methods are frowned upon in the Bayesian literature~\cite{Sivia}. Meanwhile,
Trottier {\it et al.} \cite{Trottier:2001vj} have---among other applications---employed Bayesian methods to perform extrapolations of the static-quark self energy as a function of the lattice size $L$. The extrapolant in this problem can be computed using lattice effective field theory and Trottier et al. used ``constrained curve fitting" to stabilize the fit of certain coefficients in that EFT expansion, thereby incorporating some prior information regarding the naturalness of coefficients in
the EFT expansion in their fit. However, hadronic observables, such as the behavior of the nucleon mass as a function of $m$, were not considered. 

This paper seeks to develop a general strategy for parameter estimation in EFTs: one based only on the expectation that these parameters are ``natural" with respect to the underlying scale. We believe that such priors provide a more stable and
reliable estimation of coefficients than standard fitting techniques.  This conclusion seems very general, and should apply
to a wide variety of EFT situations. We discuss those conclusions in
Sec.~\ref{Sec:Conclusion} and give a sampling of possible
applications.

\section{Bayesian probability theory}\label{Sec:Bayes}

In this section we outline the way in which we will use data to estimate the parameters in an EFT.
After a brief review of the standard maximum-likelihood technique we describe the basics of Bayesian probability theory and explain how it can be used for this problem.
In particular we show how prior information on an EFT's low-energy constants can be systematically included in the data analysis, and how the impact of higher-order effects on the extracted LECs can be accounted for by marginalization. 

Throughout this section we denote a set of data on a particular observable by $D=\{(d_k,\sigma_k):k=1,\ldots,N\}$, with $d_k$ an individual measurement at point $x_k$ and $\sigma_k$ the corresponding uncertainty.
The functional form which we want to use to describe the observable is given by $f(x,\as)$, where $f(x,\as)$ depends on a set of EFT parameters $\as=\{a_0,\ldots,a_M\}$ which we wish to determine from the data set $D$.
In any EFT application (and in many others too!) $f(x,{\bf a})$ need not and should not be assumed to be the correct functional form for the observable for all $x$. Instead we only assume that there is some $x$ domain, $x < \rho$ say, where $f$ can be systematically improved via the addition of more terms (and hence more parameters). $\rho$ would then be the breakdown scale of the EFT expansion represented by $f$.

\subsection{Maximum likelihood}

The maximum-likelihood method is often used to determine the
unknown parameters $\as$. In this method one tries to find those values of the parameters that maximize the probability of generating the data set $D$, assuming that $f(x,\as)$ is indeed the true theory, i.e. we seek to find $\as=\as_0$ such that
\begin{equation}\label{Bayes:Likelihood}
\pr(D|\as_0,f)
\end{equation}
is maximum, where 
$$
\pr(X|Y)
$$
denotes the probability of $X$ \emph{given} $Y$.
(Here and below the notation $\pr(D|\as,f)$ is used to specify both the functional form $f(x,\as)$ as well as the particular values of $\as$.)
The maximum-likelihood method simplifies further if the data are independent and the noise due to measurement is Gaussian.
In this case the probability of finding the data given the underlying functional form $f(x,\as)$ can be written as 
\begin{equation}\label{Bayes:MaxLike}
\pr(D|\as,f)=\prod_{k=1}^N\left(\frac{1}{\sqrt{2\pi}\sigma_k}\right) \exp\left(-\frac{\chi^2}{2}\right),
\end{equation}
where
\begin{equation}
\chi^2= \sum_{k=1}^N \left( \frac{d_k-f(x_k,\as)}{\sigma_k} \right)^2.
\end{equation}
Finding the maximum of $\pr(D|\as,f)$ is  equivalent to minimizing $\chi^2$, giving justification to the widely used method of least-squares.

In fact, the least-squares likelihood of Eq.~(\ref{Bayes:MaxLike}) is the least biased pdf in the case that the means $d_k$ and variances $\sigma_k$ of the $N$ uncorrelated measurements are known.
This statement can be proven via the ``principle of maximum entropy'' \cite{Jaynes:1957} which states that the least-biased pdf is found by maximizing
\begin{equation}\label{Bayes:MaxEnt}
S=-\int dx \, \pr(x)\log\left[\frac{\pr(x)}{m(x)}\right]
\end{equation} 
under the constraints of the available information, where $m(x)$ is a `measure' for the maximum entropy calculation (see App.~\ref{Sec:MaxEnt}).

\subsection{Bayesian approach}

The above approach, while standard, has three shortcomings as far as the particular application we have in mind is concerned.
The first is that maximizing Eq.~(\ref{Bayes:Likelihood}) is not exactly the problem we want to solve.
It assumes that the theory, in particular the values of the parameters $\as$, is given, while in fact the data are given and we want to infer values for $\as$. 
Mathematically we are really interested in $\pr(\as|D,f)$ (provided we assume a particular functional form $f$ to describe the data).
The question of which of these two probabilities we should try to maximize is related to a long-lasting discussion about the correct interpretation of probability, and we do not wish to comment on this here (see e.g.~\cite{Jaynes:2003,Cousins:1995,Howie:2002}).
As we will see below, the two probabilities are related by Bayes' theorem. 
The second issue is of a more practical nature.
The maximum-likelihood approach assumes no prior knowledge of the values of the parameters $\as$. 
While there are circumstances in which this is indeed appropriate, there are other cases in which information on the parameters $\as$ is available before the data analysis.
This information could come from the naturalness arguments mentioned above, from symmetry arguments, or via constraints from other experimental data. Incorporating such knowledge into ${\rm pr}({\bf a}|D,f)$ refines the EFT parameter estimate obtained from the data set $D$, and is very easy within the framework of Bayesian statistics \cite{Sivia}. Finally, the entire discussion thus far assumes a particular functional form for $f$. A more general approach would allow the extraction of relevant LECs from data using different EFT forms, for example obtained by carrying the EFT calculation to different orders in the $m/\Lambda$ expansion. Such marginalization is straightforward once ${\rm pr}(D|{\bf a},f)$ is in hand. It amounts to standard manipulations of conditional probabilities (see e.g.~\cite{Sivia}).

Bayes' theorem relates the probability of a certain parameter set being correct given a set of data $D$, $\pr(\as|D,f)$, to the probability $\pr(D|\as,f)$ of obtaining the data $D$ given the theory $f(x,\as)$ with a specific $\as$,
\begin{equation}\label{Bayes:Theorem}
\pr(\as|D,f)=\frac{\pr(D|\as,f)\pr(\as|f)}{\pr(D|f)}.
\end{equation}
Here, $\pr(\as|D,f)$ is referred to as the posterior pdf, while $\pr(\as|f)$ is called the prior pdf and incorporates information on the parameters $\as$ that we have prior to analysis of the data. 
$\pr(D|f)$ is the probability to find the data $D$ regardless of the specific values of the $a_i$ and can often be absorbed in a normalization constant, yielding
\begin{equation}
\pr(\as|D,f) \propto \pr(D|\as,f)\pr(\as|f).
\end{equation}
Here we will take the $a_i$'s to be dimensionless. In that case they are so-called ``location" parameters, and if there is no prior information on $\as$ then $\pr(\as|f)$ should be taken to be a constant~\cite{Sivia}. Consequently one finds that 
\begin{equation}
\pr(\as|D,f) \propto \pr(D|\as,f),
\end{equation}
which leads us back to the method of maximum likelihood described in the previous section.

However, in the case that prior information on the parameters \emph{is} available a constant $\pr(\as|f)$ is not appropriate and one has to decide how to incorporate the available information in the prior pdf.
This is not always straightforward, as can be seen by the example of naturalness that is of interest to us here.
In that case the information we want to encode in $\pr(\as|f)$ is that the parameters $\as$ are supposed to be natural, i.e.~of order ${\cal O}(1)$.
However, neither of these statements gives us much guidance as to what form to choose for $\pr(\as|f)$.
For instance, one way to incorporate the fact that $a_i\sim {\cal O}(1)$ would be to assign a uniform prior in the region $-5 \le a_i \le 5$:
\begin{equation}
{\rm pr}(a_i|f)=\left\{ \begin{array}{cc}
\frac{1}{10} & -5 \le a_i \le 5 \\
0 & \mbox{otherwise}
\end{array}\right..
\end{equation}
However, this is a very strict prior outside the range $[-5,5]$, and while LECs with magnitude larger than $5$ might not be ideal for the convergence of the EFT it is not clear that they should be rejected entirely.

We will incorporate naturalness in a less restrictive form.
We choose the function $f(x,\as)$ to contain $M+1$ parameters $a_i$, and we assume only that $f$ is linear in these parameters,\footnote{With our definition observables are always linear in the LECs. However, for the standard definition of the term LEC this no longer holds. The techniques developed here can be extended to non-linear dependence on the parameters but that makes the analysis more complicated, and so we defer that case to a future study.} i.e. we write
\begin{equation}
f(x,\as)=\sum_{j=0}^M a_j f_j(x),
\end{equation}
where the $f_j(x)$ are basis functions, e.g.~monomials of order $j$. We will refer to $M$ as the order of the EFT calculation. So as to simplify the notation we replace e.g. ${\rm pr}(D|{\bf a},f)$ by ${\rm pr}(D|{\bf a},M)$, since specifying the order $M$ usually defines $f$ in a given EFT. 
We then interpret the naturalness assumption as a constraint on the ensemble average of the sum of squares of the coefficients~\cite{Gull:1988}:
\begin{equation}\label{Bayes:EnsembleAve}
\left\langle\sum_{j=0}^{M} a_j^2\right\rangle =\int\! d\as\,\as^2\,\pr(\as|M,R) =(M+1)R^2,
\end{equation}
where $R$ encodes our interpretation of what ``${\cal O}(1)$" means.
We want to find the least informed pdf that incorporates the information in Eq.~(\ref{Bayes:EnsembleAve}). 
As discussed above this can be achieved by the application of the maximum entropy principle.
Using the constraint of Eq.~(\ref{Bayes:EnsembleAve}) we arrive at the prior pdf (see App.~\ref{Sec:MaxEnt})
\begin{equation}\label{Bayes:prior}
\pr(\as|M,R)=\left(\frac{1}{\sqrt{2\pi}R} \right)^{M+1}\exp \left(-\frac{\as^2}{2R^2}\right),
\end{equation}
which is a multivariate Gaussian distribution with mean $\mathbf{\mu}=\mathbf{0}$ and standard deviation $R$ and can thus be written as
\begin{equation}
\pr(\as|M,R) = \left(\prod_{i=0}^M\frac{1}{\sqrt{2\pi}R}\right)\, \exp\left(-\frac{\chi_{prior}^2}{2}\right),
\end{equation}
with
\begin{equation}
\chi_{prior}^2=\sum_{i=0}^M\frac{a_i^2}{R^2}.
\end{equation}
Note that since our testable information (\ref{Bayes:EnsembleAve}) did not include  any statement about correlations between the LECs we have obtained a $\chi_{prior}^2$ in which the $a_i$'s are uncorrelated. If correlations between different coefficients are known to exist then they should (and can) be part of the testable information provided to the maximum-entropy principle. 

Combined with a Gaussian likelihood function $\pr(D|\as,f)$ the probability of finding a theory $f(x,\as)$ given the data $D$ again assumes a Gaussian form,
\begin{equation}\label{Bayes:GaussPosterior}
\pr(\as|D,M,R)\propto \exp\left(-\frac{\chi_{aug}^2}{2}\right),
\end{equation}
where we follow Ref.~\cite{Lepage:2001ym} and introduce an ``augmented $\chi^2$'',
\begin{equation}\label{Bayes:chiAug}
\chi^2_{aug}= \chi^2+\chi_{prior}^2.
\end{equation}
The expectation values of the parameters $a_i$ are then determined by finding the maximum of the probability $\pr(\as|D,M,R)$---which is equivalent to a least-squares problem with $\chi^2$ replaced by $\chi_{aug}^2$.
This form has the advantage that techniques developed for the standard least-squares approach can be adopted, and certain manipulations can be performed analytically, as we will now demonstrate.

The standard $\chi^2$ can be written in matrix form as 
\begin{equation}\label{Bayes:MatrixChi}
\chi^2=\as^T A \as -2 \bs\cdot \as +C,
\end{equation}
where the $(M+1)\times(M+1)$ matrix $A$ is defined as
\begin{equation}
A_{ij}= \sum_{k=1}^N \frac{1}{\sigma_k^2}\, f_i(x_k)f_j(x_k),\quad i,j=0,\ldots,M,
\end{equation}
the $(M+1)$-component vector $\bs$ is given by
\begin{equation}\label{Bayes:bDef}
b_i= \sum_{k=1}^N \frac{1}{\sigma_k^2}\, d_k f_i(x_k)  , \quad i=0,\ldots,M,
\end{equation}
and
\begin{equation}\label{Bayes:CDef}
C=\sum_{k=1}^N \frac{1}{\sigma_k^2}\, d_k^2.
\end{equation}
The augmented $\chi^2$ can be written in similar form by simply replacing the matrix $A$ by $A_{aug}$,
\begin{equation}\label{Bayes:AAugDef}
A_{aug}=A+\frac{1}{R^2}\, \mathcal{I} ,
\end{equation}
where $\mathcal{I}$ is the $(M+1)\times(M+1)$ identity matrix.
The minimum of $\chi_{aug}^2$ is then simply given by
\begin{equation}
\as_0=A_{aug}^{-1}b.
\end{equation}

From these formulae it is easy to see how the size of $R$ influences the result of the regression.
As long as $1/R^2 \ll \lambda_{min}$, where $\lambda_{min}$ is the smallest eigenvalue of $A$, it will have only very little effect on the extraction of $\as_0$.
In this case the constraint by the prior information is very weak.
However, for the case that $1/R^2$ is much larger than the smaller eigenvalues of $A$, i.e.~$R$ is small, the prior information of Eq.~(\ref{Bayes:EnsembleAve}) amounts to a strong constraint on the allowed parameter values and will dominate the solution $\as_0$. The augmented $\chi^2$ is thus of most use when $R^2 \sim 1/\lambda_{min}$. The naturalness constraint can then help to refine and distinguish between what would otherwise be shallow and/or equivalent minima in the $\chi^2$ hypersurface.

\subsection{Marginalization}\label{Sec:Bayes:Marg}

In general the theory underlying the data can depend on a large number of parameters.
When we are only interested in a subset of these parameters the other (``nuisance'') parameters can be eliminated from the analysis by marginalization.

Suppose that the theory depends on the parameter sets $X$ and $Y$, where $X$ stands for the parameters of interest while $Y$ denotes all nuisance parameters.
Standard probability theory tells us that the pdf $\pr(X|D)$ can be obtained by summing/integrating the probability $\pr(X,Y|D)$ over all possible values of $Y$,
\begin{equation}\label{Bayes:Marg}
\pr(X|D)=\int dY \, \pr(X,Y|D).
\end{equation}
It is interesting to note a similarity between marginalization and effective field theory.
In both cases one ``integrates out'' those degrees of freedom one is not explicitly interested in ($Y$ and heavy degrees of freedom, respectively) and takes their contributions into account implicitly. As we shall now see,
in the case of linear dependence of $f$ on the parameters the similarity is particularly striking as the marginalization involves a Gaussian integral over the irrelevant degrees of freedom. 

Once the marginalized pdf is obtained, the problem of estimating the parameters $X$ is reduced to a lower dimensionality, thereby reducing the numerical cost. Finding the marginalized pdf has its own cost; however, for the case of a Gaussian posterior the marginalization integral can be performed analytically.
We will be interested in the case where $X$ stands for a subset of low-order parameters $\ares=(a_0,\ldots,a_{r-1})$, and $Y$ denotes higher-order parameters $\amarg=(a_r,\ldots,a_M)$ with $\as=(\ares,\amarg)$.
We want to obtain the marginalized pdf $\pr(\ares|D,M,R)$, which is given by marginalization of the posterior of Eq.~(\ref{Bayes:GaussPosterior}) over $\amarg$,
\begin{equation}\label{Bayes:GaussMargDef}
\pr(\ares|D,M,R) \propto \int d\amarg \, \exp\left(-\frac{1}{2}\,\chi_{aug}^2\right).
\end{equation}
As explained in the previous subsection we can write $\chi_{aug}^2$ as
\begin{equation}\label{Bayes:ChiAugDef}
\chi_{aug}^2=\as^T A_{aug} \as -2 \bs\cdot \as +C.
\end{equation}
Performing the integration over the parameters $\amarg$ one again obtains a Gaussian pdf,
\begin{equation}
\pr(\ares|D,M,R) \propto \exp\left[-\frac{1}{2}\left( \ares \Gamma \ares- 2\beta\cdot\ares+C \right) \right],
\end{equation}
where $\Gamma$ and $\beta$ are related to $A_{aug}$ and $\bs$ by
\begin{align}
\Gamma&=A_1-A_2(A_4)^{-1}A_3,\\
\beta&=\bres-\bmarg (A_4)^{-1}A_3,
\end{align}
where 
$\bs=(\bres,\bmarg)$ and
\begin{equation}
A_{aug}=\left(\begin{matrix}
A_1 & A_2  \\
A_3 & A_4 
\end{matrix}\right),
\end{equation}
with $A_1$ an $r\times r$, $A_2$ an $r \times (M+1-r)$, $A_3$ an $(M+1-r)\times r$, and $A_4$ an $(M+1-r)\times (M+1-r)$ matrix, respectively.
The estimates for the parameters $\ares$ are now given by
\begin{equation}
{\ares}_{,0}=\Gamma^{-1}\beta.
\end{equation}
A straightforward calculation (see App.~\ref{Sec:MargOvera}) shows that these results for ${\ares}_{,0}$ are identical to the ones obtained from the non-marginalized posterior $\pr(\as|D,M,R)$. The first $r \times r$ entries in the covariance matrix are also unaffected. Therefore marginalization has no effect on the parameter estimates in the case that the posterior pdf is Gaussian. But, for a general posterior pdf, the estimates of the parameters $\ares$ after marginalization can differ from the ones obtained from the unmarginalized pdf.

The marginalized probability $\pr(\ares|M,R,D)$ still depends on $M$ and $R$, i.e.~the choice as to which order of the EFT expansion is used to obtain the fitting function and the meaning of what is really ``natural" for the $a_i$'s.
Neither the exact order of the polynomial from which we estimate $\ares$ nor the exact value of $R$ are of specific interest to us.
Ultimately we are only interested in what the data can tell us about the value of $\ares$, and the pdf of interest is really $\pr(\ares|D)$, i.e.~we want to eliminate $M$ and $R$.
We now show how the probability $\pr(\ares|D)$ can be obtained from the familiar likelihood $\pr(D|\as,M)$ by marginalization and Bayes' theorem (also see Ref.~\cite{Gull:1988} for marginalization over $R$).

To construct $\pr(\ares|D)$ we marginalize $M$ and $R$ over suitable domains:
\begin{equation}\label{Ex:Final:Start}
\pr(\ares|D)=\sum_{M=r}^{M_{max}} \int d R \; \pr(\ares,M,R|D).
\end{equation}
Using Bayes' theorem we can rewrite the right-hand side as
\begin{equation}\label{Ex:Final:Bayes}
\sum_{M=r}^{M_{max}} \int d R \; \frac{\pr(D|\ares,M,R)\pr(\ares,M,R)}{\pr(D)}.
\end{equation}
In an $M$th-order calculation with $M > r-1$ there are additional parameters in the EFT function $f$. Thus, to calculate $\pr(D|\ares,M,R)$, we introduce them by marginalization:
\begin{align}\label{Ex:Final:amarg}
\pr(D|\ares,M,R)&=\int d\amarg \; \pr(D,\amarg|\ares,M,R) \notag \\
&=\int d\amarg \; \pr(D|\ares,\amarg,M,R)\pr(\amarg|\ares,M,R).
\end{align}
Inserting Eq.~(\ref{Ex:Final:amarg}) in Eq.~(\ref{Ex:Final:Bayes}) we find
\begin{equation}\label{Ex:Final:combined}
\pr(\ares|D)=\sum_{M}\int d R \int d\amarg \; \frac{\pr(D|\as,M,R)\pr(\amarg|\ares,M,R)\pr(\ares,M,R)}{\pr(D)}.
\end{equation}
The first probability in the numerator should be independent of our choice of $R$, i.e. $\pr(D|\as,M,R)=\pr(D|\as,M)$, and the last two terms in the numerator of Eq.~(\ref{Ex:Final:combined}) can be rewritten as
\begin{eqnarray*}
\lefteqn{\pr(\amarg|\ares,M,R) \pr(\ares,M,R)}  \\
& &= \pr(\amarg|\ares,M,R)\pr(\ares|M,R)\pr(M,R) \\
& &=\pr(\as|M,R)\pr(M,R) \\
& &=\pr(\as|M,R)\pr(M)\pr(R),
\end{eqnarray*}
where in the last step we have used that $M$ and $R$ should be independent of each other.
This gives as our final pdf
\begin{equation}\label{Ex:Final:pdf}
\pr(\ares|D)=\sum_{M=r}^{M_{max}} \int_{R_{min}}^{R_{max}} dR \int d\amarg \; \frac{\pr(D|\as,M)\pr(\as|M,R)\pr(M)\pr(R)}{\pr(D)}.
\end{equation}
Eq.~(\ref{Ex:Final:pdf}) is a key result of this paper. Its derivation employs only Bayes' theorem and the standard rules of probability. It thus encodes, in a completely general way, an EFT fitting strategy that accounts for systematic differences in fits due to results obtained with different EFT orders. It can also incorporate the requirement that EFT parameters be natural. Furthermore, the integrals over $\amarg$ that appear tend to reduce the impact on the pdf of data points where higher-order terms in the EFT are large, and so Eq.~(\ref{Ex:Final:pdf}) includes the notion of ``theoretical uncertainty" in the fitting procedure in a well-defined way. 

While Eq.~(\ref{Ex:Final:pdf}) is general the manner in which the naturalness requirement is implemented is open to interpretation. For the rest of this paper we will use the maximum-entropy prior (\ref{Bayes:prior}) for our analyses. Priors in $M$ and $R$ also need to be specified.
We will let the sum over $M$ run from $r$ to some $M_{max}$. In general one should try to ensure that the parameter estimates for $\ares$ are not sensitive to $M_{max}$. (Technically this is an implementation of an ``improper prior" on $M$ via a limiting procedure.) 
We do not have any information that would lead us to favor one value of $M$ over the other, and we therefore assign a uniform prior to $M$,
\begin{equation}\label{Ex:Final:priorM}
\pr(M)=\frac{1}{M_{max}-M_{min}+1}.
\end{equation}

Similar reasoning applies to the prior on $R$. We integrate over some region $R_{min}\le R \le R_{max}$. If the data analysis is being done in a sensible choice of units we would expect that values of $R$ close to 1 will be favored, but we do not wish to bias the fit unduly in this regard, and so we choose a uniform prior. However, since $R$ is a scale parameter the prior should be uniform in $\log(R)$ \cite{Jeffreys:1939}, not $R$, and we thus obtain
\begin{equation}\label{Ex:Final:priorR}
\pr(R)=\frac{1}{R}.
\end{equation}

Therefore while $\pr(\as|D,M,R)$ is Gaussian, the final pdf of Eq.~(\ref{Ex:Final:pdf}) after marginalization over $M$ and $R$ is no longer of Gaussian form.
This means that---unlike the case defined by Eqs.~(\ref{Bayes:GaussMargDef}) and (\ref{Bayes:ChiAugDef})---estimates for $\ares$ cannot be determined from a simple matrix multiplication.
Instead, given the pdf $\pr(\ares|D)$ we calculate the expectation values and variances of the parameters $\ares$, according to:
\begin{align}
\langle a_i \rangle &= \int d\ares\, a_i\, \pr(\ares|D), \label{Ex:ExpVal}\\
\sigma_{a_i} &=\langle a_i^2 \rangle -\langle a_i \rangle^2,
\end{align}
where we have assumed $\pr(\ares|D)$ to be normalized.

This reveals another advantage of the Bayesian approach: in addition to the uncertainty in the data, the variance $\sigma_{a_i}$ includes the uncertainties due to fitting at different $M$'s and choosing different $R$'s. These effects are included in the final pdf $\pr(\ares|D)$. Indeed, it is useful to think of Eq.~(\ref{Ex:ExpVal}) as a weighted sum over the possible values of $M$, where the weight is given by the probability $\pr(M|D)$ (c.f. Ref.~\cite{Gull:1988}). 
If one specific value of $M$ is much more likely than any other, the pdf $\pr(\ares|D)$ is dominated by this specific term in the sum and our approach will yield approximately the same answers as a fit solely at that particular order. This is in contrast to much of the EFT literature where assumptions about $M$ are often implicitly made in parameter estimation. Equation~(\ref{Ex:Final:pdf}) forces and allows such assumptions to be explicitly included in the extraction of LECs from data. 
A similar argument holds for $R$: if the pdf $\pr(R|D)$ has a spike near a particular value of $R$ then the integral over $R$ will be dominated by that value, and a fit with $R$ fixed would be quite successful. The advantage of Eq.~(\ref{Ex:Final:pdf}) is that it makes no assumptions about whether special cases associated with such peaks in the $M$ and $R$ pdfs are realized or not. Instead marginalization lets the data (together with the minimal assumptions encoded in our priors) determine which values of $R$ and $M$ will be important in the extraction of the $a_i$'s.

\section{Application to a toy problem}\label{Sec:Toy}

In the following we consider an example that allows us to illustrate the main features and advantages of Bayesian methods in fitting data in order to extract EFT parameters.
Instead of real data from an actual experiment we choose to generate artificial data from the function
\begin{equation}\label{Ex:Func}
g(x)=\left(\frac{1}{2}+\tan\left(\frac{\pi}{2} x\right) \right)^2
\end{equation}
for $x \ge 0$.
While we are not aware of any physical quantity that is described by $g(x)$, it exhibits several features that commonly appear in the analysis of data relevant to EFTs. 
The function $g(x)$ is nonanalytic for $x \,\epsilon\, \mathbb{R}$, but within a finite radius of convergence $\rho$ it can be approximated to arbitrary precision by a power series. With the application to EFTs in mind we think of $\rho$ as the ``high-energy'' scale. Therefore coefficients in the expansion of $g$ in powers of $x$  will be natural when written in units of $\rho$. 
Because of the particular argument we have chosen for the tangent function in Eq.~(\ref{Ex:Func}) the radius of convergence of the Taylor series for $g(x)$ is $\rho=1$, which simplifies the subsequent discussion. It has the consequence that the absolute values of the coefficients in the power series expansion of $g(x)$,
\begin{equation}\label{Ex:FuncExp}
g(x) \approx 0.25+ 1.57 x + 2.47 x^2 + 1.29 x^3 + 4.06 x^4 + \cdots,
\end{equation}
are ``natural" for at least the first 10 terms, with an rms value of around 3. However it is interesting to note that the coefficients do {\it not} decrease with increasing order, and so priors based on the expectation that they do~\cite{Young77} could potentially lead to misleading results.

\begin{figure}
\begin{center}
\includegraphics{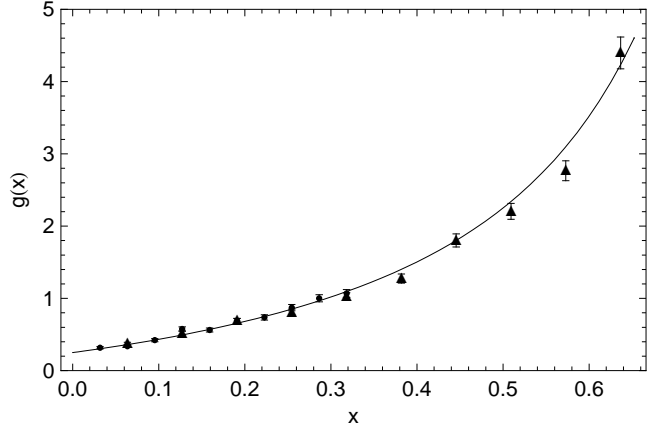}
\end{center}
\caption{Generated artificial data $D_1$ ($x_{max}=1/\pi$, circles) and $D_2$ ($x_{max}=2/\pi$, triangles). For both data sets $c=5\%$. The solid line is the function $g(x)$.\label{Ex:dataMax05Err5}}
\end{figure}

We generate ``data'' that are normally distributed about the curve $g(x)$ and we assign a relative error $c$ at each value of $x$. The data are then given by
\begin{eqnarray}
y(x_i) &=& g(x_i)(1+ c \eta_i) \\
\sigma_i &=& c \, y(x_i)
\end{eqnarray}
where the $\eta_i$ are random numbers that are normally distributed with mean $\bar \eta=0$ and variance  $\sigma_\eta=1$...
For the following example we generate two data sets $D_1$ and $D_2$, the first for $0<x \le 1/\pi$ and the second for $0<x \le 2/\pi$, each containing 10 data points with $c=0.05$.
The data are shown in Fig.~\ref{Ex:dataMax05Err5} and can be found in App.~\ref{Sec:Data}

\begin{figure}
\begin{center}
\includegraphics{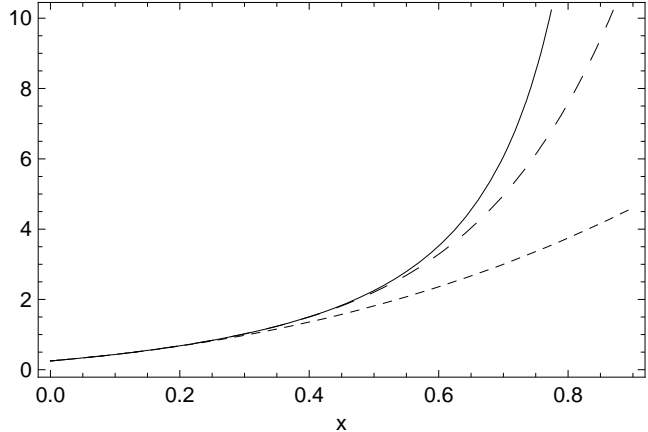}
\end{center}
\caption{The function $g(x)$ (solid line) and its power series expansion at order $M=3$ (short-dashed line) and $M=7$ (long-dashed line).\label{Ex:FullVsExp}}
\end{figure}

Our aim is to extract the coefficients of a polynomial $f_M(x,\as)$ of degree $M$ from a fit to the data, where
$$
f_M(x,\as)=\sum_{j=0}^M a_j x^j.
$$ 
As can be seen in Fig.~\ref{Ex:FullVsExp}, the power series expansion of $g(x)$ up to order 3 does not reproduce the complete function for $x\gtrsim 0.4$, while at order 7 good agreement is found up to $x \approx 0.6$.

In this section we explore three methods for extracting the coefficients $a_0$ and $a_1$ from the two different data sets shown in Fig.~\ref{Ex:dataMax05Err5}. First, we review the results obtained via a standard maximum-likelihood fit at different orders. Then we examine the way in which the augmented $\chi^2$ obtained above can be used to improve those results. Finally, we show how marginalization over $M$ and $R$ retains the improvement seen due to the use of the $\chi^2_{aug}$, and deals with the sometimes awkward issue of which values should be chosen for those two parameters.

\subsection{Standard maximum-likelihood approach}

We begin with a standard maximum-likelihood analysis of the data. Since our data are normally distributed, this reduces to a $\chi^2$-minimization problem as explained above.
The first issue is the choice of the order of the polynomial.
For the first data set with $x \le 1/\pi$ a polynomial of low order might result in values of the coefficients close to the ones in Eq.~(\ref{Ex:FuncExp}). But it seems clear from Fig.~\ref{Ex:FullVsExp} that for data corresponding to larger values of x a fit at low (e.g. third) order will not give accurate results for the true coefficients in the power series of $g(x)$.
Conversely too high an order leads to an over-constrained fit.
If, as is often the case, the function $g(x)$ is not known we will not know what order in $M$ is necessary in order to obtain reasonable likelihoods.
So here we perform the analysis at several orders, $M=1,\ldots,7$.
We start with the data set $D_1$ for which $0<x\le 1/\pi$.
The results for the first few coefficients at each order together with the corresponding $\chi^2$ per degree of freedom are given in Table~\ref{Ex:StandResultD1}.
\begin{table}[hb]
\begin{center}
\begin{tabular}{c|c|c|c|c}
$M$ & $\chi^2/d.o.f.$ & $a_0$ & $a_1$ & $a_2$  \\ \hline
1 & 2.24 & 0.203 $\pm$ 0.014 & 2.55  $\pm$ 0.11  & \\
2 & 1.64 & 0.250 $\pm$ 0.023 & 1.57  $\pm$ 0.40  & 3.33 $\pm$ 1.31  \\ 
3 & 1.85 & 0.269 $\pm$ 0.039 & 0.954 $\pm$ 1.094 & 8.16 $\pm$ 8.05 \\ 
4 & 1.96 & 0.333 $\pm$ 0.067 & -1.88 $\pm$ 2.69  & 44.7 $\pm$ 32.6 \\ 
5 & 1.39 & 0.566 $\pm$ 0.132 & -14.8 $\pm$ 6.85  & 276  $\pm$ 117 \\ 
6 & 1.85 & 0.590 $\pm$ 0.291 & -16.4 $\pm$ 18.1  & 311  $\pm$ 395 \\ 
7 & 2.67 & 0.242 $\pm$ 0.788 & 8.97  $\pm$ 56.3  & -373 $\pm$ 1494 \\ 
\end{tabular}\caption{Fit results for standard $\chi^2$ approach with $x_{max}=1/\pi$ and $c=0.05$.\label{Ex:StandResultD1}}
\end{center}
\end{table}
Surprisingly, the quadratic fit reproduces the underlying values of $a_0$ and $a_1$ extremely well. It also has the lowest $\chi^2$, and so there is a good argument for accepting this as the true value of the fit. However, if one did not know the underlying values of $a_0$ and $a_1$ one might be hard put to explain the extent to which the fit at order 2 is superior to that at order 3, or indeed, that at order 5. 
The lack of convergence for $a_0$ and $a_1$ is rather disturbing. 

This problem is exacerbated when we repeat the $\chi^2$ analysis with our  second data set $D_2$, for which $0 < x \le 2/\pi$. The results are given in Table~\ref{Ex:StandResultD2}.
\begin{table}
\begin{center}
\begin{tabular}{c|c|c|c|c}
$M$ & $\chi^2/d.o.f.$ & $a_0$ & $a_1$ & $a_2$  \\ \hline
2 & 5.35 & 0.392   $\pm$ 0.033 & -0.387 $\pm$ 0.351  & 8.08  $\pm$ 0.689 \\ 
3 & 1.47 & 0.141   $\pm$ 0.058 & 4.32   $\pm$ 0.946  & -12.7 $\pm$ 3.9 \\ 
4 & 1.48 & 0.246   $\pm$ 0.106 & 1.79   $\pm$ 2.35   & 4.81  $\pm$ 15.4 \\ 
5 & 1.46 & 0.00697 $\pm$ 0.217 & 8.67   $\pm$ 5.94   & -59.3 $\pm$ 53.2 \\ 
6 & 0.46 & 0.995   $\pm$ 0.516 & -24.0  $\pm$ 16.6   & 319   $\pm$ 187 \\ 
7 & 0.50 & 0.180   $\pm$ 1.41  & 5.98   $\pm$ 51.0   & -89.9 $\pm$ 685 \\ 
\end{tabular}\caption{Fit results for standard $\chi^2$ approach with $x_{max}=2/\pi$ and $c=0.05$.\label{Ex:StandResultD2}}
\end{center}
\end{table}
As expected, the low-order fits do not yield results in agreement with the true coefficients in the power-series expansion of $g(x)$.
But, even at higher orders, the values for the first few parameters are not close to the underlying values and have large errors. 
With knowledge of the ``true" results we can see that the $M=4$ result is the closest, but there is no minimum in $\chi^2/d.o.f.$ at order 4, and even the zeroth-order coefficient $a_0$ does not reach a stable value as the order of the fit is increased.
This is partly because $a_1$ and $a_2$ tend to much larger values than the expected natural size in this example. The best fit appears to arise from large cancellations between successive terms in the polynomial, and so standard $\chi^2$ is powerless to obtain useful information on $a_0$ and $a_1$ from the data set $D_2$. 

There are possible remedies for this problem, e.g. analyzing a subset of the data corresponding only to small $x$, but these require judgement on the part of the practitioner. How small is small enough for a finite-order polynomial to be a reasonable approximant to the underlying function? Such judgements also clearly require a trade off between making $x$ small enough that the polynomial is accurate and increasing $x$ to add more data and so increase the statistical power.

\subsection{Bayesian approach at fixed $M$ and $R$}\label{Sec:BayesianApp} 

We will now show how a Bayesian analysis of these two data sets dramatically improves the parameter estimation. In particular, we will show how the requirement of naturalness stabilizes the fit.

First we re-analyze the low-$x$ data at fixed $M$ and $R$, i.e. employ the augmented $\chi^2$ derived in Eq.~(\ref{Bayes:chiAug}). 
As explained in Sec.~\ref{Sec:Bayes} we use the prior of Eq.~(\ref{Bayes:prior}),
\begin{equation*}
\pr(\as|M,R)=\left(\frac{1}{\sqrt{2\pi}R} \right)^{M+1}\exp \left(-\frac{\as^2}{2R^2}\right),
\end{equation*}
which is a constraint not on individual parameters, but on the ensemble average: $\left\langle\as^2\right\rangle$.\footnote{Note that here, in contrast to Eq.~(\ref{Ex:ExpVal}), the integration is performed over all $\as$.} 
In this case we choose $R=1$ and perform fits at $M=2,\ldots,7$ using the data set $D_1$.
The results for the leading parameters and the corresponding probability are given in Table~\ref{Ex:BayesResultD1}.
\begin{table}
\begin{center}
\begin{tabular}{c|c|c|c|c}
$M$ & $\log[\pr(\langle\as\rangle|D_1,M,R)]$ & $a_0$ & $a_1$ & $a_2$ \\ \hline
2 & 12.00 & 0.228 $\pm$ 0.018 & 2.06 $\pm$ 0.25 & 1.60 $\pm$ 0.78 \\
3 & 11.25 & 0.230 $\pm$ 0.018 & 2.04 $\pm$ 0.25 & 1.50 $\pm$ 0.79 \\
4 & 10.35 & 0.230 $\pm$ 0.018 & 2.04 $\pm$ 0.25 & 1.49 $\pm$ 0.80 \\
5 & 9.43  & 0.230 $\pm$ 0.018 & 2.04 $\pm$ 0.25 & 1.49 $\pm$ 0.80 \\
6 & 8.51  & 0.230 $\pm$ 0.018 & 2.04 $\pm$ 0.25 & 1.49 $\pm$ 0.80 \\
7 & 7.60  & 0.230 $\pm$ 0.018 & 2.04 $\pm$ 0.25 & 1.49 $\pm$ 0.80 \\
\end{tabular}\caption{Fit results for Bayesian approach with $R=1$, $x_{max}=1/\pi$ and $c=0.05$.\label{Ex:BayesResultD1}}
\end{center}
\end{table}
One immediately sees the influence of the prior on the results. 
The estimates of the first two parameters hardly change as $M$ is increased and are much closer to the ``true'' values than in the standard $\chi^2$ approach, with the exception of the excellent quadratic fit. It is also noticeable that the uncertainty on $a_1$ and $a_2$ has decreased dramatically.
However, none of the extracted parameters lies within 1$\sigma$ of its underlying value. 
Eq.~(\ref{Ex:FuncExp}) shows that, with the exception of $a_0$, the first few coefficients in the expansion are in fact all larger than 1.
This means that the naive choice $R=1$ to express what we mean by ``natural'' is too restrictive, forcing our augmented $\chi^2$ to a local minimum that is not related to the true values of the parameters. 
As $R$ is increased, this modification of the $\chi^2$ hypersurface by the augmentation becomes less severe, allowing the fit to explore a larger domain of parameter space.
The results for $R=5$ demonstrate this, and are shown in Table~\ref{Ex:BayesResultD1R5}.
\begin{table}
\begin{center}
\begin{tabular}{c|c|c|c|c}
$M$ & $\log[\pr(\langle\as\rangle|D_1,M,R)]$ & $a_0$ & $a_1$ & $a_2$ \\ \hline
2 & 9.62   & 0.248 $\pm$ 0.023 & 1.63 $\pm$ 0.39 & 3.15 $\pm$ 1.27 \\
3 & 7.10   & 0.247 $\pm$ 0.024 & 1.65 $\pm$ 0.45 & 2.98 $\pm$ 2.32 \\
4 & 4.57   & 0.247 $\pm$ 0.024 & 1.64 $\pm$ 0.46 & 2.98 $\pm$ 2.39 \\
5 & 2.04   & 0.247 $\pm$ 0.024 & 1.64 $\pm$ 0.46 & 2.98 $\pm$ 2.39 \\
6 & -0.488 & 0.247 $\pm$ 0.024 & 1.64 $\pm$ 0.46 & 2.98 $\pm$ 2.39 \\
7 & -3.02  & 0.247 $\pm$ 0.024 & 1.64 $\pm$ 0.46 & 2.98 $\pm$ 2.39 \\
\end{tabular}\caption{Fit results for Bayesian approach with $R=5$, $x_{max}=1/\pi$ and $c=0.05$.}\label{Ex:BayesResultD1R5}
\end{center}
\end{table}
Again, all three parameters considered here show very good convergence with respect to $M$.
The central values now lie closer to the correct ones than before: all are within 1$\sigma$. 
This is in part due to the increase in the uncertainties, particularly for $a_2$, which exhibits sizeable errors.
Choosing a less restrictive prior has allowed a wider range of parameter values, while at the same time the results also tell us that the data is not sufficient to determine $a_2$ and higher-order coefficients. As expected, if we continue to increase $R$ our results go over to those of the standard $\chi^2$, so for the augmented $\chi^2$ to be a useful technique a reasonable value of $R$ must be employed.

Meanwhile Tables~\ref{Ex:BayesResultD2} and \ref{Ex:BayesResultD2R5} show the results of a similar $\chi_{aug}^2$ minimization at different orders for the data set $D_2$ for $R=1$ and $R=5$, respectively.
\begin{table}
\begin{center}
\begin{tabular}{c|c|c|c|c}
$M$ & $\log[\pr(\langle\as\rangle|D_2,M,R)]$ & $a_0$ & $a_1$ & $a_2$ \\ \hline
2 & -26.10 & 0.299 $\pm$ 0.029 & 0.807 $\pm$ 0.283 & 5.60 $\pm$ 0.55 \\
3 & -16.08 & 0.301 $\pm$ 0.029 & 1.02  $\pm$ 0.29  & 3.45 $\pm$ 0.71 \\
4 & -13.17 & 0.295 $\pm$ 0.029 & 1.16  $\pm$ 0.29  & 2.88 $\pm$ 0.74 \\
5 & -12.56 & 0.292 $\pm$ 0.029 & 1.23  $\pm$ 0.29  & 2.71 $\pm$ 0.75 \\
6 & -12.85 & 0.290 $\pm$ 0.029 & 1.26  $\pm$ 0.30  & 2.65 $\pm$ 0.75 \\
7 & -13.51 & 0.289 $\pm$ 0.029 & 1.27  $\pm$ 0.30  & 2.63 $\pm$ 0.75 \\
\end{tabular}\caption{Fit results for Bayesian approach with $R=1$, $x_{max}=2/\pi$ and $c=0.05$. \label{Ex:BayesResultD2}}
\end{center}
\end{table}
\begin{table}
\begin{center}
\begin{tabular}{c|c|c|c|c}
$M$ & $\log[\pr(\langle\as\rangle|D_2,M,R)]$ & $a_0$ & $a_1$ & $a_2$ \\ \hline
2 & -11.3 & 0.386 $\pm$ 0.033 & -0.312 $\pm$ 0.347 & 7.93  $\pm$ 0.68 \\
3 & -4.90 & 0.268 $\pm$ 0.043 & 1.96   $\pm$ 0.64  & -2.33 $\pm$ 2.52 \\
4 & -4.25 & 0.242 $\pm$ 0.045 & 2.30   $\pm$ 0.65  & -2.23 $\pm$ 2.52 \\
5 & -5.63 & 0.239 $\pm$ 0.045 & 2.28   $\pm$ 0.65  & -1.58 $\pm$ 2.56 \\
6 & -7.69 & 0.240 $\pm$ 0.045 & 2.24   $\pm$ 0.66  & -1.20 $\pm$ 2.59 \\
7 & -10.0 & 0.241 $\pm$ 0.045 & 2.21   $\pm$ 0.66  & -1.01 $\pm$ 2.61 \\
\end{tabular}\caption{Fit results for Bayesian approach with $R=5$, $x_{max}=2/\pi$ and $c=0.05$.  \label{Ex:BayesResultD2R5}}
\end{center}
\end{table}
We see that---in contrast to the unaugmented-$\chi^2$---there is no need to throw away any of the high-$x$ data in order to obtain a stable fit. The entire data set $D_2$ can be used to perform an analysis which converges with respect to the order $M$ at which the expansion is truncated. Again, the results for $a_0$ and $a_1$ obtained with $R=5$ are better than those for which $R=1$, with the results in Table~\ref{Ex:BayesResultD2R5} in agreement with the underlying values. No useful information on $a_2$ can be extracted in either case.  The results of Table~\ref{Ex:BayesResultD2} and \ref{Ex:BayesResultD2R5} also show that this data set has less power to determine the LECs $a_0$ and $a_1$ than does the lower-$x$ data set of the same statistical weight, $D_1$. 

The reader might object that this procedure is guaranteed to lead to a good fit, as an increase in the number of parameters automatically means a better fit to the data. The Tables above show that this is not the case: the logarithm of the maximum probability at fixed order $M$ peaks at a particular value of $M$ (which is, not surprisingly, higher for $D_2$ than for $D_1$) and does \emph{not} continue to grow with $M$. Higher-order fits are not always more probable since there is a ``phase-space penalty" in the pdf for introducing additional parameters into the fit~\cite{Gull:1988}.

The method used to obtain Tables~\ref{Ex:BayesResultD1}-\ref{Ex:BayesResultD2R5} is very close to that of Refs.~\cite{Lepage:2001ym,Morningstar:2001je}, which advocate the extraction of energies and amplitudes from lattice results for hadronic correlation functions via a technique the authors call ``constrained curve fitting".
Constrained curve fitting is also based on Bayesian probability and the inclusion of prior information to refine estimates of the parameters of interest.
There too maximum entropy results in a Gaussian prior pdf, which in turn gives a Gaussian posterior with a modified $\chi^2$. And, as in the case of this Section, the underlying function considered in Refs.~\cite{Lepage:2001ym,Morningstar:2001je} and Ref.~\cite{Trottier:2001vj} is an infinite sum of terms, and to perform the fit it has to be truncated at a certain order, and then convergence with respect to that truncation is sought. 
The only real difference between the methods lies in the prior information employed.
While both formulations lead to Gaussian prior pdfs, Refs.~\cite{Lepage:2001ym,Morningstar:2001je,Trottier:2001vj} employ information on the mean and variance of the individual parameters, while in this paper only the less restrictive knowledge of an ensemble average is assumed.
One would arrive at the prior pdf Eq.~(\ref{Bayes:prior}) in the formalism of Refs.~\cite{Lepage:2001ym,Morningstar:2001je,Trottier:2001vj} by assigning $\mu=0$ and $\sigma=R$ to each individual parameter. Therefore our method can be considered a special case of constrained curve fitting.
However, if more detailed information on individual parameters is available, it can and should be employed.

The results of this section show that, as might be expected, the choice of $R$ has an impact on parameter estimation.
If the true values of parameters are not known, even the notion of naturalness does not give clear guidance on which $R$ to choose, as both $R=1$ and $R=5$ are natural.
We therefore advocate marginalization over $R$ over a suitable region, thereby taking into account all reasonable values of $R$.

\subsection{Bayesian approach including marginalization over $M$ and $R$}

In the fits of the previous section the value of $a_2$ is not well constrained by data set $D_2$, and neither data set is sufficient to extract useful information about the coefficients of the cubic (and higher) terms. Therefore from now on $a_2$, $a_3$, \ldots will be considered nuisance parameters. This will allow us to focus on the pdf for $a_0$ and $a_1$.
We adopt $\ares=(a_0,a_1)$, and compute the
pdf $\pr(\ares|D)$ using Eq.~(\ref{Ex:Final:pdf}). That pdf depends on the choice of the range of $M$ values over which we marginalize. Above we advocated increasing $M_{max}$ until convergence with respect to that parameter was obtained. Convergence for this example (both $D_1$ and $D_2$) is obtained by $M_{max}=8$, with the results for $a_0$ and $a_1$ unaffected by the inclusion of higher orders beyond that in the marginalization over $M$. Such a result was to be expected given the peaks in the posterior probability as a function of $M$ seen in Tables \ref{Ex:BayesResultD1}--\ref{Ex:BayesResultD2R5}, and the fact that the results for $\langle a_0 \rangle$ and $\langle a_1 \rangle$ presented there are not altered as $M$ is increased from $5$ to $6$ to $7$ and so on. 

With this in mind we choose $M_{min}=2$, $M_{max}=8$, $R_{min}=0.1$, $R_{max}=10$, and compute the pdf of Eq.~(\ref{Ex:Final:pdf}) using data set $D_1$. The results for $a_0$ and $a_1$ are
\begin{align}
a_0 &= 0.239 \pm 0.021 \label{Ex:a0result1}\\
a_1 &= 1.84 \pm 0.37. \label{Ex:a1result1}
\end{align}
The results here are consistent with those shown in Tables~\ref{Ex:BayesResultD1} and \ref{Ex:BayesResultD1R5}, but we emphasize that (\ref{Ex:a0result1}) and (\ref{Ex:a1result1}) include not only marginalization over $M$ but also marginalization over $R$. Since the bulk of the contributions to the $R$ integral come from $R$ between 2 and 5 and the errors are quite $R$ dependent the errors found in Eqs.~(\ref{Ex:a0result1}) and (\ref{Ex:a1result1}) are a little larger than those in Table~\ref{Ex:BayesResultD1}, but are smaller than those in Table \ref{Ex:BayesResultD1R5}. Marginalizing over $R$ thus incorporates the systematic uncertainty due to the ambiguity in which value of $R$ to choose, and does so in a way that lets the data determine which values of $R$ should dominate the marginalization integral. 

As a check on the stability of this procedure we have repeated the calculation choosing $R_{max}=20$, resulting in
\begin{align*}
a_0 &= 0.239 \pm 0.022 \\
a_1 &= 1.83 \pm 0.37,
\end{align*}
which is consistent with (\ref{Ex:a0result1}) and (\ref{Ex:a1result1}).  While it might violate the strict principles of a Bayesian analysis, examining the results for dependence on $R_{max}$ and looking for a plateau like the one seen here serves as a way to check that one has made a safe choice for $R_{max}$. (Note that there is no corresponding dependence on $R_{min}$, since the extremely restrictive priors corresponding to small $R$ contribute little to the final pdf.)

Obviously in most real applications the underlying values of the parameters are not known {\it a priori}. But, in this case we have the "true" values of $0.25$ and $\approx 1.57$ in hand, and we see that the Bayesian result is a much better extraction of the values of $a_0$ and $a_1$ than the ones obtained from a standard $\chi^2$ analysis of this data set (with the exception of the surprisingly good quadratic fit).  This builds confidence in the marginalization strategy represented by Eq.~(\ref{Ex:Final:pdf}).

Applying Eq.~(\ref{Ex:Final:pdf}) to data set $D_2$, again with $M_{min}=2$, $M_{max}=8$, $R_{min}=0.1$, $R_{max}=10$, we obtain
\begin{align}
a_0 &= 0.241 \pm 0.048 \label{Ex:a0result3}\\
a_1 &= 2.23 \pm 0.74. \label{Ex:a0error3}
\end{align}
These are again consistent with results obtained by minimizing $\chi^2_{aug}$ and seeking convergence with $M$ (as long as the choice of $R$ is not too restrictive). Eqs.~(\ref{Ex:a0result3}) and (\ref{Ex:a0error3}) reproduce the underlying coefficients in the Taylor series much better than does a standard $\chi^2$ analysis of $D_2$. 
The errors are larger than in the case of the $D_1$ results, since the data extends to larger $x$, and so higher-order effects are more important. 
Another indication that the data set $D_2$ does not determine the parameters as well as the set $D_1$ comes from the observation of small residual $R_{max}$-dependence of $a_1$: with
With $R_{max}=20$ we obtain
\begin{align}
a_0 &= 0.237 \pm 0.050\\
a_1 &= 2.28 \pm 0.80,
\end{align}
while $R_{max}=40$ gives
\begin{align}
a_0 &= 0.237 \pm 0.051\\
a_1 &= 2.30 \pm 0.82.
\end{align}
While $a_0$ does not depend on $R_{max}$, the dependence of $a_1$ and its error on this choice shows that the data in set $D_2$ cannot determine the central value to better than two digits, or the standard deviation to better than one digit. 

We have also extracted the pdf based on only a subset of the data set $D_2$.
We chose its first 5 data points, which span the range $0 < x \le 1/\pi$, and denote this by $\overline{D}_2$. 
(This is \emph{not} the same data as in set $D_1$, which contains 10 data points.)
The results for $M_{min}=2$, $M_{max}=8$ and $R_{min}=0.1$, $R_{max}=10$ are
\begin{align}
a_0 &= 0.238 \pm 0.038 \label{Ex:a0result4}\\
a_1 &= 2.12 \pm 0.49.
\end{align}

We consider it a strength of our technique that we can analyze either a subset or the full data set and obtain consistent results. Comparison of the analyses using $D_2$ and $\overline{D}_2$ shows that $a_0$ and $a_1$ are mainly determined by $\overline{D}_2$. (This conclusion is also apparent from the fact that the standard deviation for $a_0$ in Eq.~(\ref{Ex:a0result3}) is significantly larger than the naive $c/\sqrt{N}$.) But the Bayesian techniques allow us to use {\it all} the available data without prejudice as to which $x$'s are `too large' for an EFT calculation at a given order $M$ to be reliable. Indeed, data on intervals  $[0,x_{max}]$ with $x_{max}$ quite close to $\rho$ can be employed for parameter estimation. One might be concerned that this will lead to the use of data with $x > \rho$, but, at least in the example considered in this section, including such data in the fit leads to a ridiculously small posterior pdf $\pr(\ares|D)$. The probability of the parameters given the data goes through an abrupt drop due to the singularity that cannot be reproduced by the ``EFT" polynomial form. 

Thus Bayesian methods should allow EFT practitioners to perform reliable parameter estimation without undue concern about whether data is safely within the region of validity of the EFT. If data that is beyond the reach of the EFT is employed for the fitting the behavior of the posterior pdf $\pr(\ares|D)$ will provide signals of the theory's breakdown. An important caveat is, however, that in many examples (e.g. that of Sec.~\ref{Sec:Model}) the physical quantity in question will not have a pole on the real axis when the radius of convergence is reached, and so the drop in $\pr(\ares|D)$ will not be as dramatic as it is here.

\section{Nucleon mass in chiral perturbation theory}\label{Sec:Nucleonmass}

Having demonstrated the key features of our Bayesian approach using a toy model in the previous section, we now turn to an application in an actual EFT, chiral perturbation theory. In particular we will explore the chiral expansion of the nucleon mass in SU(2) $\chi$PT, which can be written as (see e.g.~\cite{Gasser:1987rb,McGovern:1998tm,Becher:1999he,Schindler:2007dr})
\begin{align}\label{Mass:Exp}
M_{\chi PT}(m) &= M_0+k_1 m^2 + k_2 m^3 + k_3 m^4 \log\left(\frac{m}{\mu}\right)+k_4 m^4+k_5 m^5 \log\left(\frac{m}{\mu}\right) \notag\\
&\quad +k_6 m^5+k_7 m^6 \log\left(\frac{m}{\mu}\right)^2 + k_8 m^6\log\left(\frac{m}{\mu}\right) +k_9 m^6+ {\cal O}(m^7).
\end{align}
Here $M_0$ is the nucleon mass in the chiral limit and $m$ denotes the lowest-order pion mass.
The coefficients $k_i$ in Eq.~(\ref{Mass:Exp}) are linear combinations of the coefficients of the operators that appear in the $\chi$PT Lagrangian. Specific expressions for the LECs $k_i$ in terms of these coefficients can be found in Ref.~\cite{Schindler:2007dr} and we do not reproduce them here. As stated in the Introduction it will be the $k_i$'s that we focus on fitting, specifically $M_0$, the nucleon mass in the chiral limit, and $k_1$, which is proportional to the nucleon sigma term. 

With the notation of Eq.~(\ref{Mass:Exp}) the LECs $k_i$ are dimensionful quantities, with increasing inverse powers of some energy scale.
We now rewrite Eq.~(\ref{Mass:Exp}) to make that scale explicit:
\begin{align}\label{Mass:ExpRes}
\frac{M_{\chi PT}(m)}{\Lambda} &= \frac{M_0}{\Lambda}+ \frac{\tilde k_1}{\Lambda^2} m^2 + \frac{\tilde k_2}{\Lambda^3}m^3 + \frac{\tilde k_3}{\Lambda^4}m^4 \log\left(\frac{m}{\mu}\right)+ \frac{\tilde k_4}{\Lambda^4}m^4+  \frac{\tilde k_5}{\Lambda^5}m^5 \log\left(\frac{m}{\mu}\right) \notag\\
&\quad + \frac{\tilde k_6}{\Lambda^5}m^5 + \frac{\tilde k_7}{\Lambda^6}m^6 \log\left(\frac{m}{\mu}\right)^2 +  \frac{\tilde k_8}{\Lambda^6}m^6\log\left(\frac{m}{\mu}\right) + \frac{\tilde k_9}{\Lambda^6}m^6+\ldots,
\end{align}
where we have also rescaled the nucleon mass since here we restrict ourselves to theories with only one high-energy scale. A critical question for any attempt to use Eq.~(\ref{Ex:Final:pdf}) to analyze data on the behavior of $M$ as a function of $m$ is now: for what scale $\Lambda$ are the dimensionless $\tilde{k}_i$'s of Eq.~(\ref{Mass:ExpRes}) natural?

One might expect that they would all be natural with respect to the nominal breakdown scale of $\chi$PT, $\Lambda=4 \pi F$, with $F$ the pion decay constant. But several of the $k_i$'s are significantly larger than is indicated by naive-dimensional analysis with respect to this scale. For example, the coefficient $k_5$, which defines the leading non-analytic contribution at the two-loop level, is given entirely in terms of low-order coefficients:
\begin{equation}
k_5 = \frac{3\texttt{g}_A^2}{1024\pi^3F^4}(16\texttt{g}_A^2-3)=\frac{1}{(4 \pi F)^4}\frac{3 \pi \texttt{g}_A^2 (16 \texttt{g}_A^2 - 3)}{4}.
\label{Mass:k5}
\end{equation}
Strictly, $\texttt{g}_A$, the axial coupling of the nucleon, and $F$ both take their chiral-limit values here. But the difference between physical values and chiral-limit values is a higher-than-fifth-order effect, and using physical values for evaluation we obtain
$\tilde{k}_5$ of roughly 85, if $\Lambda=4 \pi F$.

A second example is the coefficient $k_4$, which is often written as
\begin{equation}
k_4 = -\hat{e}_1-\frac{3}{128\pi^2 F^2 M_0}(\texttt{g}_A-c_2 M_0),
\end{equation}
where $c_2$ is a coefficient in the second-order pion-nucleon Lagrangian, ${\cal L}_{\pi N}^{(2)}$, and 
the contribution $\hat{e}_1$ stems from ${\cal L}_{\pi N}^{(4)}$.\footnote{There does not seem to be a consistent notation for this coefficient, as it is also referred to as $e_1^{(4)}$ or simply $e$, and sometimes also defined with the opposite sign.} A fit to lattice data on $M_N(m)$ using the fifth-order $\chi$PT form  obtained $\hat{e}_1=-30.5$ GeV$^{-3}$,\footnote{The actual value of $\hat{e}_1$ depends on the choice of renormalization scale $\mu$, which in Ref.~\cite{McGovern:2006fm} was taken to be $\mu=m_{\pi}^{\rm{phys}}$. Here we are not concerned with its actual value, but just its approximate magnitude, which is only logarithmically dependent on $\mu$.} which is perhaps not surprising given a $k_5$ of $47.56$ GeV$^{-4}$~\cite{McGovern:2006fm}. 

Here we choose $\hat{e}_1=-8$ GeV$^{-3}$ which is comparable in magnitude to the value extracted in Ref.~\cite{McGovern:2006fm}. (This value is also not unreasonable given that $\hat{e}_1=16 e_{38} + 2 e_{115} + 2 e_{116}$ in terms of the coefficients in the Lagrangian of Ref.~\cite{Fettes:2000gb}, if we assume that the operators in that ${\cal L}_{\pi N}^{(4)}$ are the ones whose coefficients are ${\cal O}(1)$.) This, together with the physical values of $\texttt{g}_A$ and $F$ and the values of the LECs $c_1, c_2,c_3$ as found in Ref.~\cite{Becher:2001hv}, as well as the mesonic LECs from Ref.~\cite{Gasser:1983yg} and $d_{16}=-1.93\,\text{GeV}^{-2}$ \cite{Fettes:1999wp} (all evaluated at the scale $\mu=M_0$) yields:
\begin{equation}
k_1=3.84, k_2=-5.63, k_3=6.49, k_4=8.28, 
k_5=47.56,  k_6=70.53, k_7=12.8,
\label{Mass:Consts}
\end{equation}
where each $k_i$ is expressed in appropriate units of (GeV)$^{-n}$. 
The values of $k_8$ and $k_9$ depend on a number of unknown higher-order LECs, and for the demonstrative purposes of this section we set them to 
\begin{equation}\label{Mass:Consts2}
k_8=10, k_9=-100.
\end{equation}
so that when we supplement $M_{\chi PT}(m)$ by a model at higher $m$ in the next section we obtain a smooth function for $M_N(m)$ over a range of $m$ up to 1 GeV.

The value we choose for $M_0$ is 0.88 GeV, but this particular choice is not relevant to the success or failure of our parameter-estimation strategy, since we will extract it from our pseudo-data using Eq.~(\ref{Ex:Final:pdf}). This pseudo-data is not data from lattice QCD, but instead consists of 11 data points produced using the formula (\ref{Mass:Exp}) and the constants (\ref{Mass:Consts}) in the region between $m=200$ and $m=500\,\text{MeV}$. We added normally-distributed offsets of magnitude 1.5\% to the underlying form (\ref{Mass:Exp}).This error is quite conservative given the precision reached in modern lattice calculations (see, e.g., the compilation in Appendix~B of Ref.~\cite{Musch:2005si}).

In the following we take the basis functions in Eq.~(\ref{Ex:Final:pdf}) to be 
\begin{equation}
\begin{array}{lllll}
f_0(x)=1; & f_2(x)=x^2; & f_3(x)=x^3; & f_{4a}(x)=x^4 \log(x),& f_{4b}(x)=x^4;\\
f_{5a}(x)=x^5 \log(x), &f_{5b}(x)=x^5;& f_{6a}(x)=x^6 \log^2(x),& f_{6b}(x)= x^6 \log(x),& f_{6c}(x)=x^6;
\end{array}
\label{Mass:basis}
\end{equation}
with $x \equiv m/\Lambda$. Initially we choose $\Lambda=1$ GeV. We consider calculations at various different {\it chiral orders}, not with various different numbers of basis functions, e.g. the three functions $f_{6a}$, $f_{6b}$, and $f_{6c}$ are all added to the fit together when we go from order $P=5$ to order $P=6$, although each functions have an independent coefficient.  We use a modified version of Eq.~(\ref{Ex:Final:pdf}) that takes account of this distinction between chiral order and number of basis functions.
In order to demonstrate our general method we continue to use the prior of Eq.~(\ref{Bayes:prior}) in this first study of the problem. We note, though, that $\chi$PT does provide additional information on most of the coefficients of the
basis functions which are non-analytic in the quark mass (see, e.g. Eq~(\ref{Mass:k5})), and that in future studies such information could be used to refine the prior on the $k_i$'s.

\begin{figure}
\begin{center}
\includegraphics{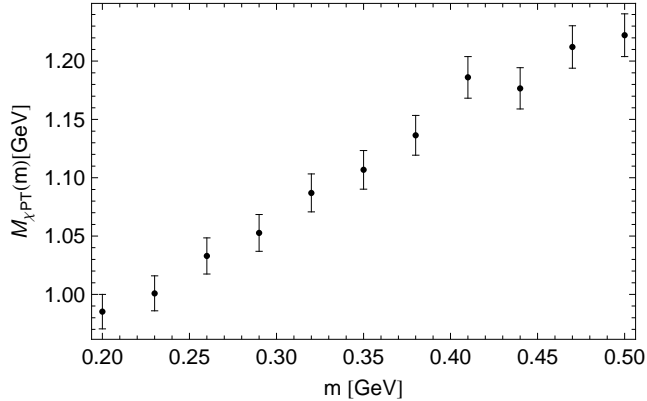}
\caption{Data generated from Eq.~(\ref{Mass:Exp}) with 1.5\% error. \label{Mass:ChiralData}}
\end{center}
\end{figure}

We focus on the first two parameters, which are $M_0$ and $k_1$. Marginalization is performed from $P_{min}=3$ to $P_{max}=6$. $R_{min}$ is chosen to be 0.1, and we vary $R_{max}$ to produce the results shown in Table~\ref{Mass:RDep1GeV}. We do not see any appreciable $R_{max}$ dependence in the results. This lack of $R_{max}$ dependence  might lead one to trust these as reliable LEC extractions, but in fact the central value for $M_0$ is slightly more than $1\sigma$ from the ``true'' value ($M_0=0.88\,\mbox{GeV}$), while the discrepancy for $k_1$ is more than 2$\sigma$ ($k_1=3.84\,\mbox{GeV}^{-1}$). 

\begin{table}
\begin{center}
\begin{tabular}{c|c|c}
$R_{max}$  & $M_0$ (GeV) & $ k_1$ (GeV$^{-1}$) \\ \hline
10  & 0.917 $\pm$ 0.027 & 1.98 $\pm$ 0.73  \\
20  & 0.916 $\pm$ 0.026 & 1.98 $\pm$ 0.74  \\
40  & 0.916 $\pm$ 0.027 & 1.98 $\pm$ 0.74  \\
80  & 0.916 $\pm$ 0.027 & 1.98 $\pm$ 0.75 \\
160 & 0.916 $\pm$ 0.027 & 1.98 $\pm$ 0.75 \\
240 & 0.917 $\pm$ 0.026 & 1.97 $\pm$ 0.74 \\
400 & 0.917 $\pm$ 0.026 & 1.98 $\pm$ 0.73 \\
\end{tabular}\caption{Dependence of estimated parameters on $R_{max}$. The fit was performed with $m=200-500\, \text{MeV}$ and $\Lambda=1\,\text{GeV}$. Marginalization is over chiral order $P=3-6$ with $R_{min}=0.1$. The data were generated with $M_0=0.88 \,\text{GeV}$ and $k_1=3.84\,\text{GeV}^{-1}$. \label{Mass:RDep1GeV}}
\end{center}
\end{table}

We attribute this failure to the fact that several of the LECs used to generate our pseudo-data are not natural with respect to the scale $\Lambda$=1\,\mbox{GeV}. As discussed above, the scale $\Lambda$ at which the low-energy constants of the EFT are natural must be specified in order to use our ``naturalness prior", i.e. the units in which the LECs are expected to be $O(1)$, must be chosen. Several of the LECs in Eqs.~(\ref{Mass:Consts}) and (\ref{Mass:Consts2}) are not ${\cal O}(1)$ for $\Lambda=1\,\mbox{GeV}$. The marginalization over $R$ is not sufficient to compensate for such misidentification of the EFT's underlying scale. While---especially for low-order coefficients---trade-offs between $\Lambda$ and $R$ are possible, a poor choice for $\Lambda$ means that higher-order coefficients grow without bound.
Changing $\Lambda$ amounts to a change of units and so makes no difference to fits using the standard $\chi^2$, but choosing $\Lambda$ too big means that higher-order coefficients make 
disproportionately large contributions to $\chi^2_{prior}$. 
Correct identification of $\Lambda$ is thus key to reliable parameter estimation using the formula (\ref{Ex:Final:pdf}). In practice this choice of $\Lambda$ then supplies part of the prior used to produce the posterior pdf. From now on we make this dependence on the underlying scale explicit by writing the posterior pdf as $\pr(\ares|D,\Lambda)$.

The literature on baryon $\chi$PT together with the observed coefficients  in the expansion (\ref{Mass:Exp}) suggest a scale $\Lambda=0.5\,\text{GeV}$ is reasonable. $\chi$PT may cease to be {\it useful} for $m$ significantly below 500 MeV, e.g. the 350 MeV identified in Refs.~\cite{McGovern:2006fm,Bernard08,Djukanovic:2006xc}, but with $\Lambda=500$ MeV as the formal radius of convergence of the $\chi$PT expansion for $M_N(m)$ the coefficients considered here, while still a bit large in some cases, are ${\cal O}(1)$:
\begin{equation}
\begin{array}{lllll}
\tilde M_0=1.76, & \tilde k_1=1.92, & \tilde k_2=-1.41, & \tilde k_3=0.81, & \tilde k_4=1.03, \\
\tilde k_5=2.97, & \tilde k_6=4.41, & \tilde k_7=0.4, & \tilde k_8=0.31, & \tilde k_9=-3.12.
\end{array}
\end{equation}

Repeating the analysis of the data displayed in Fig.~\ref{Mass:ChiralData}
with the set of basis functions (\ref{Mass:basis}), this time defined with $\Lambda=500$ MeV, we again marginalize over $P_{min}=3$ to $P_{max}=6$ with $R_{min}=0.1$. We obtain the results shown in Tab.~\ref{Mass:RDep500MeV}, where for the convenience of the reader we have converted the results back to units that are powers of a GeV.
\begin{table}
\begin{center}
\begin{tabular}{c|c|c}
$R_{max}$  & $M_0$ (GeV)& $ k_1$ (GeV$^{-1}$) \\ \hline
10  & 0.91 $\pm$ 0.03 & 2.11 $\pm$ 1.36 \\
20  & 0.91 $\pm$ 0.03 & 2.11 $\pm$ 1.35 \\
40  & 0.91 $\pm$ 0.03 & 2.13 $\pm$ 1.37 \\
80  & 0.91 $\pm$ 0.03 & 2.10 $\pm$ 1.36 \\
160 & 0.91 $\pm$ 0.03 & 2.08 $\pm$ 1.33  \\

\end{tabular}\caption{Dependence of estimated parameters on $R_{max}$. The fit was performed with $m=200-500\, \text{MeV}$ and $\Lambda=500\,\text{MeV}$. Marginalization over chiral order $P=3-6$ and $R_{min}=0.1$. \label{Mass:RDep500MeV}}
\end{center}
\end{table}

As in the case  $\Lambda=1$ GeV that is presented in Tab.~\ref{Mass:RDep1GeV}, we see only very weak $R_{max}$ dependence. Indeed, our analysis of the integrals over the parameter $R$ shows that the main contribution to the results presented here stems from the region $R \approx 1-2$. This is reassuring, since it suggests that naturalness is compatible with the data if $\Lambda$ is chosen to be 500 MeV. Also reassuring is that 
the lowest-order parameter, $M_0$, now agrees with the underlying value at the $1\sigma$ level. But the central value of $k_1$ found by fitting the pseudo-data is still a bit more than 1$\sigma$ below the underlying $k_1$, with both the error and the central vale of $k_1$ increasing noticeably compared to the values presented in Table~\ref{Mass:RDep1GeV}. 

This trend persists as $\Lambda$ is lowered further. In Tab.~\ref{Mass:LambdaDep} we show results for $M_0$ and $k_1$ at several values of $\Lambda$. Since the $R_{max}$ dependence is very small for all considered values, we only show results obtained with $R_{max}=10$. The extracted values for $M_0$ do not change for $\Lambda$ below $0.4\,\mbox{GeV}$, and agree with the underlying value there. The parameter $k_1$, on the other hand, does not show any plateau in the considered range of $\Lambda$: its central value increases monotonically as $\Lambda$ decreases. Once $\Lambda \leq 0.4$ GeV the extracted $k_1$ agrees (within the 1$\sigma$ error) with the underlying one. It is interesting to see that the error on $k_1$ is largest at $\Lambda=0.4\,\mbox{GeV}$ and decreases for smaller values of $\Lambda$. However, this behaviour is peculiar to $k_1$: if one tries to extract higher-order parameters from the data their errors behave $\sim 1/\Lambda^n$, where $n$ is the dimension of the LEC in question. In the case that the underlying scale of the EFT is not well known, a study of the $\Lambda$ dependence of the extracted LECs might be a useful check to see whether the results are reliable. Viewed in that light, Table~\ref{Mass:LambdaDep} suggests that, absent specific assumptions regarding $\Lambda$, the only conclusion we can draw about $k_1$ from these data is that it is larger than 0.8 GeV$^{-1}$ and smaller than 4.1 GeV$^{-1}$.

\begin{table}
\begin{center}
\begin{tabular}{c|c|c}
$\Lambda$ (GeV)  & $M_0$ (GeV)& $ k_1$ (GeV$^{-1}$) \\ \hline
1    & 0.92 $\pm$ 0.03 & 1.98 $\pm$ 0.73 \\
0.5  & 0.91 $\pm$ 0.03 & 2.19 $\pm$ 1.36 \\
0.4  & 0.90 $\pm$ 0.04 & 2.50 $\pm$ 1.43 \\
0.35 & 0.89 $\pm$ 0.03 & 2.72 $\pm$ 1.34 \\
0.3  & 0.89 $\pm$ 0.03 & 2.88 $\pm$ 1.21 \\
0.25 & 0.89 $\pm$ 0.03 & 2.99 $\pm$ 1.09 \\
\end{tabular}\caption{Dependence of estimated parameters on the scale $\Lambda$. The fit was performed with $m=200-500\, \text{MeV}$, marginalization over chiral order $P=3-6$ and $R=0.1-10$. \label{Mass:LambdaDep}}
\end{center}
\end{table}

Since this nucleon-mass data set gives only a weak constraint on $k_1$, we now marginalize over it too. The resulting $M_0$ extraction does not show any $\Lambda$ dependence in the range $\Lambda=300-500\,\mbox{MeV}$. We find:
\begin{equation}
M_0=0.91 \pm 0.04~{\rm GeV}.
\end{equation}
This agrees with the results shown in Table~\ref{Mass:LambdaDep}, where marginalization over $k_1$ was not performed, and is consistent with the underlying value of $M_0=0.88$ GeV.

We have also performed a standard least-squares fit to the data at several orders. The results are shown in Table~\ref{Mass:StandRes}. The result with the best $\chi^2/d.o.f.$ gives an acceptable value for $M_0$, although with a much larger error than is obtained from the Bayesian approach. That result is not, though, stable with respect to the order of the fit. The standard $\chi^2$ also shows that $k_1$ cannot be usefully constrained from these data.
\begin{table}
\begin{center}
\begin{tabular}{c|c|c|c}
$P$  & $M_0$ (GeV) & $ k_1$ (GeV$^{-1}$) &  $\chi^2$/d.o.f. \\ \hline
3 & 1.12 $\pm$ 0.025 & -7.60  $\pm$ 0.63 & 4.13 \\
4 & 1.02 $\pm$ 0.29  & -16.4  $\pm$ 33.8 & 0.95 \\
5 & 4.37 $\pm$ 7.07  & -987   $\pm$ 2303 & 1.24 \\
\end{tabular}\caption{Results of standard least-square fit at different chiral orders $P$ for $\Lambda=1\,\text{GeV}$. \label{Mass:StandRes}}
\end{center}
\end{table}

\section{Fitting nucleon-mass data beyond $\chi$PT's domain of validity}

\label{Sec:Model}

The fits performed in the previous section had the advantage that the fit function included all of the terms in the underlying function used to generate our pseudo-data. In this section we examine what happens when---as in Sec.~\ref{Sec:Toy}---the EFT form breaks down at $m=\Lambda$, and goes over to some other function which {\it cannot} be written in terms of the basis functions used in the fitting procedure. This transition to non-EFT dependence of the physical quantity on the independent variable may be smooth: one does not necessarily expect a singularity in the quantity of interest as the EFT's breakdown scale is crossed (c.f. Sec.~\ref{Sec:Toy}). Nevertheless, we shall show that the tools used to analyze the toy-model data of Sec.~\ref{Sec:Toy} allow consistent fits across the boundary at $\Lambda$, and permit diagnosis of the situation in which the fit includes too much data that is outside the EFT's domain of validity.

\begin{figure}
\begin{center}
\includegraphics{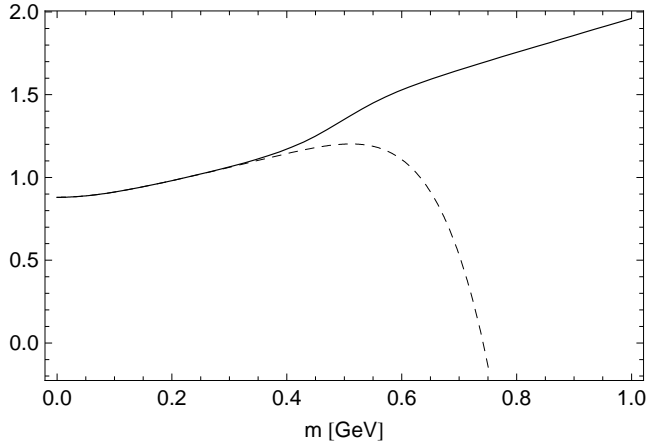}
\end{center}
\caption{The functions $M_{\chi PT} (m)$ (dashed line) and $M_N(m)$ (full line).\label{Mass:ChiralVsModel}}
\end{figure}

For our analysis we modify the function used to generate artificial nucleon-mass data by smoothly ``turning off'' the chiral dependence and ``turning on'' a model dependence for pion masses above a scale $\Lambda$:
\begin{equation}\label{Mass:smooth}
M_N(m)=M_{\chi PT} (m)\left(1 - g\left(\frac{m}{\Lambda}\right)\right) + M_{\rm model}(m) g\left(\frac{m}{\Lambda}\right),
\end{equation}
where $g$ is a smooth function obeying $g(0)=0$ and $\lim_{x \rightarrow \infty} g(x)=1$. In order to have $\chi$PT be valid below the scale $\Lambda$ but breakdown {\it at} $\Lambda$ one should choose $g(x)$ to have a Taylor-series expansion about $x=0$ with a radius of convergence of 1. In order not to disturb the terms up to sixth order  in $M_{\chi PT}(m)$ and the terms there that are non-analytic in the quark mass (odd in $m$) one should also demand that $g$ be even in $x$ and have vanishing second-, fourth- and sixth-order coefficients. These requirements are satisfied by
\begin{equation}
g(x)=\frac{2}{\pi}\arctan(x^8).
\end{equation}
In the previous section we argued for a value of 500 MeV for the underlying EFT scale $\Lambda$, and that is what we choose in the form (\ref{Mass:smooth}) here. As in Ref.~\cite{Leinweber:2000} the function $M_{\rm model}(m)$ is chosen to have the correct heavy-quark limit:
\begin{equation}\label{Mass:ModelNuc}
M_{\rm model}(m)=\alpha+\beta m.
\end{equation}
The resulting $M_N(m)$ contains {\it linear} dependence on the pion mass $m$ which is not present in the functions with which we analyze the data.\footnote{Ref.~\cite{Leinweber:2000} argues that $m_\pi^2 \sim m_q$ at quark masses below the charm-quark mass, so the form (\ref{Mass:ModelNuc}) would not become appropriate until much higher $m_q$. For our purposes the key point is that the linear-in-$m$ term is not present in our fit function $M_{\chi PT}(m)$, and so its appearance means that the $\chi$PT expansion has definitely broken down above $m=\Lambda$. See also Ref.~\cite{WalkerLoud:2008bp} where a linear fit to $M(m)$ works to surprisingly low $m_q$.}
We select parameter values
\begin{equation}
\alpha=1\,\text{GeV}; \qquad \beta=1,
\end{equation}
to give a smooth transition around $m=\Lambda$.
The resulting functions $M_{\chi PT} (m)$ and $M_N(m)$ are shown in Fig.~\ref{Mass:ChiralVsModel}.

We generate several data sets between 200 MeV and a varying $m_{max}$, each with $11$ data points. We choose $\Lambda=0.5\,\text{GeV}$ and statistical uncertainties of 1.5\%. In our analysis of the resulting pseudo-data we marginalize over $P=3$ to $P=6$ and set $R_{min}=0.1$. 

We begin by presenting results for $m_{max}=350\,\text{MeV}$, which places all data far enough below $\Lambda$ that we expect the non-chiral piece of $M_N(m)$ will not play a large role. The central values and 1$\sigma$ errors on $M$ and $k_1$ for different $R_{max}$'s are shown in Table~\ref{Mass:Model350}.
\begin{table}
\begin{center}
\begin{tabular}{c|c|c}
$R_{max}$  & $M_0$ (GeV)& $ k_1$ (GeV$^{-1}$) \\ \hline
10  & 0.91 $\pm$ 0.04 & 1.61 $\pm$ 1.86 \\
20  & 0.91 $\pm$ 0.04 & 1.61 $\pm$ 1.88 \\
40  & 0.91 $\pm$ 0.05 & 1.62 $\pm$ 1.90 \\
80  & 0.91 $\pm$ 0.04 & 1.61 $\pm$ 1.89 \\
160 & 0.91 $\pm$ 0.04 & 1.59 $\pm$ 1.85  \\
\end{tabular}\caption{Dependence of estimated parameters on $R_{max}$ for data generated with Eq.~(\ref{Mass:smooth}). The fit was performed with $m=200-350\, \text{MeV}$ and $\Lambda=500\,\text{MeV}$. Marginalization over chiral order $P=3-6$ and $R_{min}=0.1$. \label{Mass:Model350}}
\end{center}
\end{table}
As was the case in Sec.~\ref{Sec:Nucleonmass}, we find that the extracted LECs are largely independent of $R_{max}$. Their values are consistent within the errors with those obtained from fitting the ``purely chiral" data in the mass range $200~{\rm MeV} \leq m \leq 500~{\rm MeV}$.\footnote{They also agree very well with the LECs extracted from a ``purely chiral'' data set covering the pion-mass range $200~{\rm MeV} < m < 350~{\rm MeV}$.} At these low pion masses we do not gain much of a constraint on $k_1$ from a data set consisting of 11 points with 1.5\% errors.

Next, we consider the effects of the transition across $m=\Lambda$ in the fit. For this exercise we choose $R_{max}=10$ as this is in agreement with our knowledge of the the underlying values of the LECs (we have checked that our results have only minor $R_{max}$ dependence). 
Table~\ref{Mass:ModelMMax} then shows the results for several values of $m_{max}$. Note that since we employ Eq.~(\ref{Mass:smooth}) to generate the data, the high-$m$ data points start to deviate from the form $M_{\chi PT}(m)$ starting around 450 MeV, as can be seen in Fig.~\ref{Mass:ChiralVsModel}.
\begin{table}
\begin{center}
\begin{tabular}{c|c|c|c}
$m_{max}$ (MeV)  & $M_0$ (GeV) & $ k_1$ (GeV$^{-1}$) &  $\log\left[\pr(\ares|D,M,\Lambda)\right]$ \\ \hline
350  & 0.91 $\pm$ 0.04 & 1.61 $\pm$ 1.86 & 1.97 \\
400  & 0.92 $\pm$ 0.04 & 1.54 $\pm$ 1.54 & 2.31 \\
450  & 0.92 $\pm$ 0.03 & 1.35 $\pm$ 1.27 & 2.71 \\
500  & 0.93 $\pm$ 0.03 & 1.11 $\pm$ 1.07 & 3.01 \\
700  & 0.95 $\pm$ 0.04 & 0.50 $\pm$ 1.27 & 2.12 \\
1000 & 0.88 $\pm$ 0.03 & 2.43 $\pm$ 0.94 & -0.03 \\
\end{tabular}\caption{Dependence of estimated parameters on $m_{max}$ for data generated with Eq.~(\ref{Mass:smooth}). The fit was performed with $m=200-m_{max}\, \text{MeV}$ and $\Lambda=500\,\text{MeV}$. Marginalization over chiral order $P=3-6$ and $R=0.1-10$. \label{Mass:ModelMMax}}
\end{center}
\end{table}
The results show a drop in the probability once data sets with $m_{max}$ larger than 500 MeV are considered. Indeed, the maximum probability obtained with the data set that extends to $m_{max}=1\,\mbox{GeV}$ is more than on order of magnitude smaller than that found in the case of $m_{max}=500\,\mbox{MeV}$. $\chi$PT is, unsurprisingly, less likely to be the theory that describes data sets which extend beyond 500 MeV. 

As expected, $k_1$ is only weakly constrained by any of the fits that have reasonable values of $\log\left[\pr(\ares|D,M,\Lambda)\right]$. 
The results for $M_0$ are relatively stable even when a sizable part of the data is generated with $m>\Lambda$, although in a few cases the underlying value is outside the $1\sigma$ range. 

Marginalization over $k_1$ does not improve the situation. Results of extractions of $M_0$ alone for data sets with different $m_{max}$ are shown in Tab.~\ref{Mass:ModelmMaxk1marg}. Once again, some of the extracted values are more than $1\sigma$ from the underlying value. We attribute these failures to reproduce the underlying value to the fact that $\Lambda=500$ MeV does not yield LECs in the chiral expansion of $M_N$ that are particularly natural. 
Investigation on how the parameter extraction changes as we vary the scale that appears in the naturalness prior is ongoing.

\begin{table}
\begin{center}
\begin{tabular}{c|c|c}
$m_{max}$ (MeV)  & $M_0$ (GeV)  &  $\log\left[\pr(\ares|D,M,\Lambda)\right]$ \\ \hline
350  & 0.91 $\pm$ 0.04 & 2.14 \\
400  & 0.92 $\pm$ 0.03 & 2.41 \\
450  & 0.92 $\pm$ 0.02 & 2.55 \\
500  & 0.92 $\pm$ 0.02 & 2.25 \\
700  & 0.94 $\pm$ 0.04 & 1.58 \\
1000 & 0.88 $\pm$ 0.03 & 2.34 \\
\end{tabular}\caption{Dependence of estimated parameters on $m_{max}$ for data generated with Eq.~~(\ref{Mass:smooth}). The fit was performed with $m=200-m_{max}\, \text{MeV}$ and $\Lambda=500\,\text{MeV}$. Marginalization over chiral order $P=2-6$ and $R=0.1-10$. \label{Mass:ModelmMaxk1marg}}
\end{center}
\end{table}

For the fit with $m_{max}=700$ MeV five of the eleven data points are above $\Lambda$. 
In order to examine what happens if we fit a subset of these data we have used the first 6 data points in that set (those with $m < \Lambda$) and found: 
\begin{align*}
M_0 &= 0.93 \pm 0.04~{\rm GeV}, \\
k_1 &= 1.17 \pm 1.38~{\rm GeV}^{-1}.
\end{align*}
The result is consistent with that obtained using the full $m_{max}=700$ MeV data set, indicating that the fit to those data is driven by the lower-mass points.

We have used different functional forms to model the high-$m$ behavior, such as negative values for the slope in the linear model of Eq.~(\ref{Mass:ModelNuc}) and a quadratic dependence on the pion mass. The results of these investigations are qualitatively the same as those presented here.

We therefore conclude that the formula (\ref{Ex:Final:pdf}) can be used, together with a $\chi$PT fit form, to obtain the pdf $\pr(\ares|D,\Lambda)$ from data on $M_N(m)$ that deviates from the prediction of $\chi$PT above $m=\Lambda$. Fits of data over different intervals in $m$---including intervals extending beyond $\Lambda$---yield consistent results for $M_0$ and $k_1$. When the fit extends to $m > \Lambda$ the result is driven by data that lie in the chiral regime unless such high $m$'s are considered that the fit becomes exceedingly unlikely. However, if the results extracted of $M_0$ and $k_1$ are to be in agreement with the underlying values of these parameters, a value of $\Lambda$ that corresponds to LECs of $O(1)$ must be used in the application of the nautralness prior. 

\section{Conclusions}\label{Sec:Conclusion}

When used for the extraction of effective-field-theory low-energy constants from data standard methods for parameter estimation require some art in their application. Only data that are within the domain of validity of the EFT should be used for this purpose, and the process of deciding what that domain is may require iteration: the data used for the fit must be successively trimmed to yield a reasonable result for LECs. Furthermore, one of the criteria for  a ``reasonable result" for the EFT fit is that there should not be large cancellations between different orders in the EFT expansion, and that the resulting EFT LECs should be natural. 

In the Bayesian framework all of this information is put into the analysis from the beginning. Indeed a strict Bayesian would incorporate all the requirements mentioned in the previous paragraph in the prior pdf and then report only the resulting posterior pdf. The fact that this posterior pdf is precisely the quantity that those trying to obtain LECs are interested in, $\pr(\as|D,f)$, makes Bayesian methods and EFT a very good match, as does the fact that in both EFT and Bayesian approaches unwanted degrees of freedom are ``integrated out" of the calculation, leaving one free to focus on the entities that determine the low-energy dynamics. 

One straightforward way to employ Bayesian methods in EFT parameter estimation is through minimization of the augmented $\chi^2$, a procedure that has been dubbed ``constrained curve fitting" and advocated for the analysis of lattice correlation functions in Refs.~\cite{Lepage:2001ym,Morningstar:2001je}. In Secton~\ref{Sec:Bayes} we showed how a ``naturalness prior" obtained using the principle of maximum entropy leads straightforwardly to an augmented $\chi^2$. In Sec~\ref{Sec:BayesianApp} we showed, via a toy problem, that the use of the prior information stabilizes the fit of EFT LECs with respect to the order, $M$,  at which the extraction is done. It also refines the parameter estimates, resulting in smaller errors---especially on parameters beyond $a_0$.

The parameter $R$ that encodes the ambiguity in the notion of ``${\cal O}(1)$" coefficients is an input to the naturalness prior. An approach that is more general than constrained curve fitting focuses, not on the pdf $\pr(\ares|D,R,M)$ that was computed in Sec.~\ref{Sec:BayesianApp}, but on $\pr(\ares|D)$. This requires marginalization over $M$ and $R$, which can be straightforwardly done using standard rules of probability and Bayes' theorem, yielding Eq.~(\ref{Ex:Final:pdf}):
\begin{displaymath}
\pr(\ares|D)=\sum_{M=r}^{M_{max}} \int_{R_{min}}^{R_{max}} dR \int d\amarg \; \frac{\pr(D|\as,M)\pr(\as|M,R)\pr(M)\pr(R)}{\pr(D)}.
\end{displaymath}
While similar in general spirit to constrained curve fitting, analysis of an EFT problem using Eq.~(\ref{Ex:Final:pdf}) could be considered more general. Eq.~(\ref{Ex:Final:pdf}) accounts for marginalization over the order $M$, and thereby includes the uncertainty due to the truncation of the fitting function. It also requires only very weak assumptions about the value of $R$. 

We then applied these methods to the chiral expansion of the nucleon mass, $M_N$, as a function of the pion mass $m$. We analyzed pseudo-data generated from both the $\chi$PT form of $M_N$ and a form which included additional physics at $m > 500$ MeV. We did this 
by fitting the value of {\it all} coefficients in the chiral expansion up to terms of ${\cal O}(m^6)$. In both cases we found that the nucleon mass in the chiral limit, $M_0$, could be determined with reasonable precision by data in the mass range $m=200$--$350$ MeV. Using data at higher values of $m$---including values above the breakdown scale of $\chi$PT---produced results that were consistent with the fit results from this low-$m$ data set. The maximum likelihood of the fit also dropped significantly once $\chi$PT was used to extract $M_0$ and $k_1$ using data that extended well beyond the theory's domain of validity. 

The formula (\ref{Ex:Final:pdf}) with the prior (\ref{Bayes:prior}) for $\pr(\as|M,R)$ assumes dimensionless coefficients. In any practical EFT calculation a scale $\Lambda$ must be chosen to absorb the increasingly negative mass dimension of the higher-order LECs. Large (in units of GeV$^{-n}$) coefficients in the function $M_N(m)$ emphasize the importance of this choice. In general the posterior pdf obtained from Eq.~(\ref{Ex:Final:pdf}) depends on the units used once the ``naturalness prior" is employed, and so the posterior pdf should be written $\pr(\ares|D,\Lambda)$. We analyzed one set of pseudo-data on $M_N(m)$ choosing a range of $\Lambda$'s. The value of $M_0$ does not change for $\Lambda$'s below $400$ MeV. In this range the value of $k_1$ obtained from the fit was consistent (at the 1$\sigma$ level) with the underlying value of this parameter used to generate the pseudo-data. We conclude that the nucleon-mass data used here were not sufficiently accurate to determine $k_1$ unless some prior knowledge of the underlying $\chi$PT scale was supplied.

An alternative way to perform the analysis of nucleon-mass data would be to assume that the coefficients of the terms in Eq.~(\ref{Mass:Exp}) that are non-analytic in the quark mass (i.e. contain odd powers or logs of $m$) are known. The simplest way to perform that analysis would be to subtract the non-analytic pieces from the data, and fit the remaining terms:
\begin{equation}
M_{\rm analytic}(m)=M_0 + k_1 m^2 + k_4 m^4 + k_9 m^6
\end{equation}
to the resulting points. This, then, is simply a polynomial regression, where the application of Bayesian methods has been well discussed in the literature~\cite{Young77,BlightOtt75,Deaton80}. Here we instead fitted all coefficients, although we marginalize $k_n$ for $n > 1$. This also means that our results take account of the issue that some non-analytic terms in the chiral expansion of the nucleon mass are not well-constrained by other data. It would be interesting to develop priors that improve upon the ``naturalness prior" (\ref{Bayes:prior}) by incorporating what is known about these coefficients in the nucleon-mass fit.

With the groundwork for a Bayesian analysis of data on $M_N(m)$ in place it would appear very worthwhile to apply the methods developed here to actual lattice data on this quantity, see e.g. Refs.~\cite{Musch:2005si,WalkerLoud:2008bp} and references therein. Because of its ability to build in prior information on LECs, including information on their correlations, 
our method could serve as an alternative to approaches like bootstrapping of the data, as done e.g. in \cite{WalkerLoud:2008bp}. Our method could also be extended to the case of a global fit of different LECs that appear in several physical quantities, as advocated in 
Ref.~\cite{WalkerLoud:2008bp}. Such tasks are, however, beyond the scope of the present work. 

More generally, the techniques developed here can be extended to EFT expansions that depend on more than one variable and so contain non-analytic functions of the ratios of two different low-energy scales, e.g. particle energy $E$ divided by $m_\pi$. This would permit the extraction of EFT parameters from multi-energy  analyses of experimental data (see, for example, Refs.~\cite{Beane:2002wn} and \cite{Rentmeester:2003mf} for two such treatments using conventional statistical techniques). Such analyses of experimental data will, however, require treatment of the case in which observables have non-linear dependence on the underlying EFT parameters.  

Finally, we point out that Bayesian methods are also well-suited to making predictions in EFTs. Using similar steps to those that produced a formula for the marginalized pdf $\pr(\ares|D,\Lambda)$, we can calculate the pdf, $\pr(M_N(m)|D,\Lambda)$, that  predicts $M_N(m)$ at some $m$ where there is no extant lattice calculation given the existing data $D$. That pdf will be calculable 
as a sum/integral over the result found for $M_N(m)$ at different orders, calculated with different higher-order coefficients~\cite{BMAreview}. Bayesian methods therefore yield predictions for physical quantities that have a well-defined uncertainty. That uncertainty incorporates both the uncertainty due to the input data, {\it and} the uncertainty due to higher-order effects in the EFT expansion. As such we expect it to grow as the EFT expansion becomes less accurate at higher $m$. The ability to provide such information on the reliability of a theoretical calculation is one of the great benefits of EFT, and it can be fully realized using Bayesian techniques. 

\acknowledgments{MRS and DRP are grateful for encouraging and informative conversations with D.~Drabold, R.~Furnstahl and J.~McGovern, and to R.~Furnstahl for comments on the manuscript. DRP also thanks M.~Birse, C.~Morningstar, and U. van Kolck for stimulating discussions on these topics. We also thank J.~McGovern for pointing out some errors in the original version of this paper. This research was supported by the US DOE under grant no. DE-FG02-93ER40756 and by the Science and Technology Facilities Council of the United Kingdom.}

\begin{appendix}

\section{Determination of pdfs: via the principle of maximum entropy}\label{Sec:MaxEnt}

Priors incorporate previous knowledge of the hypothesis to be tested.
As explained above for the example of naturalness, it is not always clear which functional form is best suited to incorporate this knowledge in a probability density.
The method of maximum entropy \cite{Jaynes:1957} has been proposed as a way to obtain pdfs in cases where testable information is available.
Maximum entropy is a variational method that gives the least biased pdf $\pr(x)$ by maximizing the entropy
\begin{equation}\label{MaxEnt:Ent}
S=-\int dx \, \pr(x) \, \log\left[\frac{\pr(x)}{m(x)}\right]
\end{equation}
under the constraints of the previously available information.
Here, $m(x)$ is a `measure' that renders Eq.~(\ref{MaxEnt:Ent}) invariant under a change of variables.
The most basic testable information is the normalization of the pdf,
\begin{equation}\label{MaxEnt:Norm}
\int dx \, \pr(x)=1.
\end{equation}

Let us consider the case of the prior information of Eq.~(\ref{Bayes:EnsembleAve}),
\begin{displaymath}
\left\langle\sum_{j=0}^{M} a_j^2\right\rangle =(M+1)R^2.
\end{displaymath}
We then need to maximize the entropy $Q$
\begin{align}\label{MaxEnt:Q}
Q  = &-\int d\as \, \pr(\as|M,R) \, \log\left[\frac{\pr(\as|M,R)}{m(\as)}\right]+\lambda_0 \left[ 1 -\int d\as \, \pr(\as|M,R)\right] \notag\\ &+\lambda_1\left[(M+1)R^2-\int d\as\, \as^2 \, \pr(\as|M,R) \right],
\end{align}
as a functional of $\pr(\as|M,R)$ and a function of the Lagrange multipliers $\lambda_0$ and $\lambda_1$.
Here we have indicated that the prior information holds at a fixed order $M$ (i.e.~a fixed number of $a_j$'s) and known $R$.
Performing this maximization  one finds a maximum for:
\begin{equation}
\pr(\as|M,R)=m(\as) e^{-(1-\lambda_0)}e^{-\lambda_1\as^2}.
\end{equation}
Assuming a uniform measure $m(\as)=const$ then gives:
\begin{equation}\label{MaxEnt:prior}
\pr(\as|M,R)=\left(\frac{1}{\sqrt{2\pi}R} \right)^{M+1}e^{-\frac{\as^2}{2R^2}}.
\end{equation}

If both the mean $\langle a_j \rangle= a_{j,0}$ and standard deviation $\sigma_{a_j}$ of each parameter is known a similar analysis leads to a product of Gaussians for the pdf
\begin{equation}
\pr(\as|\as_0,\sigma_{\as})=\prod_{j=0}^{M}\frac{1}{\sqrt{2\pi}\sigma_{a_j}}\exp\left( -\frac{(a_j-a_{j,0})^2}{2\sigma_{a_j}^2} \right),
\end{equation}
which is the standard maximum-entropy derivation of the $\chi^2$ distribution~\cite{Sivia}. 
It should be noted that the pdf obtained from maximum entropy depends on the choice of the measure $m(x)$ of Eq.~(\ref{MaxEnt:Ent}) being constant.
A different choice of $m(x)$ leads to different results for the pdf given the same testable information.

\section{Marginalization over higher-order parameters in the linear case}\label{Sec:MargOvera}

In Sec.~\ref{Sec:Bayes:Marg} we showed how marginalization can be used to eliminate nuisance parameters from our considerations.
For fixed order $M$ and fixed $R$ we obtained a posterior pdf $\pr(\ares|D,M,R)$.\footnote{Here we consider the general case $\ares=(a_0,\ldots,a_{r-1})$.}
According to Eqs.~(\ref{Bayes:GaussMargDef}) and (\ref{Bayes:ChiAugDef}) we can write this posterior in the form
$$
\pr(\ares|D,M,R) \propto \int d\amarg \, \exp\left(-\frac{1}{2}\,\chi_{aug}^2\right),
$$
where
$$
\chi_{aug}^2=\as^T A_{aug} \as -2 \bs\cdot \as +C
$$
for the linear case, and $A_{aug}$, $b$ and $C$ are given in Eqs.~(\ref{Bayes:AAugDef}), (\ref{Bayes:bDef}), and (\ref{Bayes:CDef}) respectively.
We now show that marginalization over $\amarg$ does not change the result for the expectation value of $\ares$ and its variance.

To perform the integration over the $\amarg$ in Eq.~(\ref{Bayes:GaussMargDef}) we rewrite $\bs$ as $\bs=(\bres,\bmarg)$ such that
\begin{equation}
\bs\cdot\as=\bres\cdot\ares+\bmarg\cdot\amarg.
\end{equation}
Analogously we write the matrix $A_{aug}$ in block form,
\begin{equation}\label{Marg:Block}
A_{aug}=\left(\begin{matrix}
A_1 & A_2  \\
A_3 & A_4 
\end{matrix}\right),
\end{equation}
with $A_1$ an $r\times r$, $A_2$ an $r \times (M+1-r)$, $A_3$ an $(M+1-r)\times r$, and $A_4$ an $(M+1-r)\times (M+1-r)$ matrix, respectively.
We can then write $\chi_{aug}^2$ as
\begin{equation}
\chi_{aug}^2= \ares A_1 \ares -2\bres\cdot\ares + \amarg A_4 \amarg -2(\bmarg-\ares A_3)\cdot\amarg,
\end{equation}
where we have used that $A_2^T=A_3$.
The integration over $\amarg$ is now straightforward and we obtain for the marginalized posterior pdf
\begin{align}
\pr(\ares|D,M,R)&=\left(\prod_i \frac{1}{\sqrt{2\pi}\sigma_i} \right)\left(\prod_j \frac{1}{\sqrt{2\pi}\sigma_{p_j}}\right) \sqrt{\frac{(2\pi)^{M+1-r}}{\det A_4}} \notag \\ 
& \quad \times \exp\left[-\frac{1}{2}\left( \ares \Gamma \ares- 2\beta\cdot\ares+C \right) \right],
\end{align}
where we have defined
\begin{align}
\Gamma&=A_1-A_2(A_4)^{-1}A_3, \label{Marg:GammaDef}\\
\beta&=\bres-\bmarg (A_4)^{-1}A_3,
\end{align}
and we have included the appropriate normalization factors.
The marginalized posterior is again of Gaussian form and the value of $\ares$ that maximizes $\pr(\ares|D,M,R)$ is given by
\begin{equation}\label{Marg:aresResult}
{\ares}_{,0}=\Gamma^{-1}\beta.
\end{equation}

For the considered case the marginalization procedure does not change the results for the coefficients $\ares$.
In the fit with the complete set of parameters the vector $\as$ that maximizes the probability is given by
\begin{equation}
\as_0=A_{aug}^{-1}b.
\end{equation}
To determine the first $r$ components of $\as_0$ we need to calculate $A_{aug}^{-1}$. With $A_{aug}$ in block form (see Eq.~(\ref{Marg:Block})), the inverse can also be written in block form,
\begin{equation}\label{Marg:InverseBlock}
A_{aug}^{-1}=\left(\begin{matrix}
(A_1-A_2 A_4^{-1}A_3)^{-1} & -A_1^{-1}A_2(A_4-A_3A_1^{-1}A_2)^{-1}  \\
-A_4^{-1}A_3(A_1-A_2 A_4^{-1}A_3)^{-1} & (A_4-A_3A_1^{-1}A_2)^{-1} 
\end{matrix}\right),
\end{equation}
where the upper left matrix is just the inverse of $\Gamma$ defined in Eq.~(\ref{Marg:GammaDef}).
The first $r$ components of $\as_0$ are then given by
\begin{equation}
\as_{0,res}=\Gamma^{-1}\bres-A_1^{-1}A_2(A_4-A_3A_1^{-1}A_2)^{-1}\bmarg,
\end{equation}
while Eq.~(\ref{Marg:aresResult}) reads
\begin{equation}\label{Marg:aresRes}
{\as}_{res,0}=\Gamma^{-1}\bres-\Gamma^{-1}A_2A_4^{-1}\bmarg.
\end{equation}
The two results agree if 
\begin{equation}
A_1^{-1}A_2(A_4-A_3A_1^{-1}A_2)^{-1}=\Gamma^{-1}A_2A_4^{-1}.
\end{equation}
Multiplying with $A_4-A_3A_1^{-1}A_2$ from the right and $\Gamma=A_1-A_2 A_4^{-1}A_3$ from the left shows that this matrix equation indeed holds and therefore the result for the first $r$ components of $\as_0$ remain invariant under marginalization, i.e.~ $\as_{0,res}=\as_{res,0}$.
Note also that the inverse of $\Gamma$ is related to the covariance matrix for the marginalized case.
From Eq.~(\ref{Marg:InverseBlock}) one sees that the elements $\rm{(cov)}_{i,j}$ with $i,j<r$ of the full covariance matrix are the same as the elements of the covariance matrix in the marginalized case since $(A_1-A_2 A_4^{-1}A_3)^{-1}=\Gamma^{-1}$.

\newpage

\section{Data for toy model application}\label{Sec:Data}

\begin{table}[h]
\begin{center}
\begin{tabular}{l|c|c}
$\frac{\pi}{2}$ x & d(x) & $\sigma$\\ \hline
0.05 & 0.31694 & 0.01585 \\
0.1  & 0.33844 & 0.01692 \\
0.15 & 0.42142 & 0.02107 \\
0.2  & 0.57709 & 0.02885 \\
0.25 & 0.56218 & 0.02811 \\
0.3  & 0.68851 & 0.03443 \\
0.35 & 0.73625 & 0.03681 \\
0.4  & 0.87280 & 0.04364\\
0.45 & 1.0015 & 0.0501 \\
0.5  & 1.0684 & 0.0534
\end{tabular}\caption{Data set $D_1$\label{App:D1Table}}
\end{center}

\begin{center}
\begin{tabular}{l|c|c}
$\frac{\pi}{2}$  x & d(x) & $\sigma$\\ \hline
0.1 & 0.37385 & 0.01869 \\
0.2 & 0.51985 & 0.02599 \\
0.3 & 0.68911 & 0.03446 \\
0.4 & 0.81065 & 0.04053 \\
0.5 & 1.0268 & 0.0513 \\
0.6 & 1.2747 & 0.0637 \\
0.7 & 1.8016 & 0.0901 \\
0.8 & 2.2042 & 0.1102 \\
0.9 & 2.7660 & 0.1383 \\
1   & 4.3970 & 0.2198
\end{tabular}\caption{Data set $D_2$\label{App:D2Table}}
\end{center}
\end{table}

\end{appendix}

\end{document}